\documentclass[preprint]{aastex}
\usepackage{float}
\usepackage{color}
\usepackage{multirow}

\title{A CSO Broadband Spectral Line Survey of Sgr B2(N)-LMH from 260 - 286 GHz}
\author{Brett A. McGuire \& P. Brandon Carroll}
\affil{Division of Chemistry and Chemical Engineering, California Institute of Technology, Pasadena, CA 91125}
\author{Anthony J. Remijan}
\affil{National Radio Astronomy Observatory, Charlottesville, VA 22903}
\email{aremijan@nrao.edu}

\begin{document}

\begin{abstract}

Presented here are the results of a broadband spectral line survey of the Sgr B2(N) - LMH region from 260 - 286 GHz using the Caltech Submillimeter Observatory.  The data were taken over the course of a single night (May 26, 2013) during the course of science testing of the remote observational capabilities of the facility.  The data are freely available to public both as raw, double-side band observational data and as a minimally-reduced ascii spectrum.  The procedural scripts used for the preliminary data reduction using CLASS are provided as well.  The observational parameters and preliminary data reduction procedures are detailed.  Finally, we provide instructions for accessing the data as well as comment on the robustness of the preliminary reduction.

\end{abstract}

\section{Introduction}
\label{intro}

Advances in the state-of-the-art during the last decade have made the acquisition of broadband, high-resolution, and high-sensitivity spectral line surveys a time-efficient process.  These surveys have great staying power. By collecting data over a wide range in frequency space, rather than focusing on small windows around molecular transitions of interest, the data can be mined repeatedly for new molecular information as laboratory spectra become available. Perhaps the most studied target of these surveys is the galactic center source Sagittarius B2(N) (see Table \ref{surveys}).

\begin{deluxetable}{c c c}
\tablewidth{0pt}
\tablecaption{Summary of spectral line surveys of Sgr B2(N)}
\tablecolumns{3}
\tablehead{ \colhead{Covered Frequencies} &  \colhead{Facility} & \colhead{Reference}}
\startdata
0.3 - 50 GHz 				&		GBT							&	(see, e.g. McGuire et al. 2012)			\\
\\
80 - 116 GHz				& 		IRAM 30-m					&	(Belloche et al. 2008)				\\
\\
86 - 90 GHz				&	\multirow{2}{*}{BIMA \& NRAO 12-m}	&	\multirow{2}{*}{\citep{Friedel2004}}		\\
106 - 110 GHz																						\\
\\
140 - 170 GHz				&		NRAO 12-m					& 	(B.E. Turner 1991)\tablenotemark{a}		\\
\\
202 - 218 GHz				&	IRAM 30-m						&	(Belloche et al. 2008)				\\
\\
218 - 265 GHz				&		SEST						&	\citep{Nummelin1998}  				\\
\\
260 - 286 GHz				&		CSO							&	(This work)						\\
\\
480 - 1910 GHz			&		Herschel Space Telescope		&	\citep{Neill2012}					\\
\enddata
\tablenotetext{a}{Details of the observations and motivations can be found in Remijan et al. (2008)}
\tablecomments{GBT - Robert C. Byrd 100-m Green Bank Telescope; IRAM - Institut de Radioastronomique Millim\'{e}trique; BIMA - Berkeley-Illinois-Maryland Array; SEST - Swedish-ESO Submillimeter Telescope; CSO - Caltech Submillimeter Observatory}
\label{surveys}
\end{deluxetable}

This work has been performed in the spirit of the \textbf{PR}ebiotic \textbf{I}nterstellar \textbf{MO}lecular \textbf{S}urvey (PRIMOS) project.  The PRIMOS key project began in January of 2008 and observations continue to expand its frequency coverage.  This project provides high-resolution, high-sensitivity spectra of the Sgr B2(N-LMH) complex centered at (J2000) $\alpha$ = 17h47m110s, $\delta$ = -28$^{\circ}$22\arcmin17$\arcsec$.  The PRIMOS project is providing reduced data to the public with no proprietary period.  Here, we endeavor to the do the same.

\clearpage

\section{Observations}
\label{obs}

The data presented here towards of the Sgr B2(N-LMH) complex are centered at (J2000) $\alpha$ = 17h47m110s, $\delta$ = -28$^{\circ}$22\arcmin17$\arcsec$.  The observations were conducted on May 26, 2013 at the Caltech Submillimeter Observatory as part of science testing of the facility's remote observational capabilities using the 8 GHz broadband 230 GHz receiver.  Pointing solutions were acquired every 2 hours on average, with pointing offsets consistent with previous nights to within a few arcseconds.  The data were obtained in position switching mode with an offset of 2$^{\circ}$.

The 230 GHz receiver at the CSO is a double-side band (DSB) receiver.  Thus, multiple IF settings are necessary for a reliable deconvolution to single-side band (SSB) spectra.  For these observations, three rest frequencies, separated by $\sim$8 GHz, were observed with at least 5 IF settings.  This provides at least 5 different rest- and image-frequency combinations for each observed frequency.\footnote{Necessarily less redundancy is obtained near the edges of the observations.}  Several additional, deeper integrations were obtained at slightly different rest frequencies and IFs and have been included in this data set.

The data were processed by two FFTS spectrometers. The first provides 1 GHz of DSB coverage with a frequency resolution of 122 kHz.  The second FFTS spectrometer provides two, 4-GHz windows of DSB coverage with a frequency resolution of 269 kHz.

\section{Data Analysis}
\label{data}

The complete, raw dataset and reduction scripts are accessible at:\\
 \url{http://tracker.cv.nrao.edu/PRIMOS/CSO_260_290_Survery_Data/}.  

The preliminary reduction is presented here and is available in ascii format through the \textbf{S}pectral \textbf{Li}ne \textbf{S}earch \textbf{E}ngine (SLiSE) accessible at:
 \url{http://www.cv.nrao.edu/~aremijan/SLiSE/}.

The procedures used to clean, baseline subtract, and deconvolve the data for the preliminary reduction presented here are given in plain text in Appendix A and will be described here in detail.

The reduction was performed using the CLASS package of the GILDAS suite of programs.\footnote{http://www.iram.fr/IRAMFR/GILDAS}  The raw observational spectra are adjusted to a rest frequency corresponding to a $V_{lsr}$ = +64 km/s.  First, the FFTS1 data were Hanning Smoothed to $\sim$244 kHz, then both FFTS1 and FFTS2 data were resampled to a resolution of 300 kHz (0.33 km/s at 275 GHz).  Next, noise spurs from the spectrometers were removed and blanked.  Finally, several scans were dropped from the record after a visual inspection revealed inconsistencies with the rest of the data set.

Continuum and baseline subtraction for spectral line observations of Sgr B2(N) is non-trivial.  A cursory inspection of the DSB data reveals the spectra are likely confusion-limited, and virtually no baseline is visible.  For the purposes of the preliminary reduction presented here, a third-order polynomial is fit to each spectrum and removed.  This is discussed further below.

Finally, the spectra are deconvolved, allowing for intensity variations between sidebands.  Because of the broad linewidths present in Sgr B2(N), the resulting spectrum is then smoothed twice more to a final resolution of 1.2 MHz (1.3 km/s at 275 GHz).  This smoothed spectrum is presented in its entirety in Figure 1, and in 500 MHz-wide increments in Figures 2 - 53.  The unsmoothed data have an RMS of $\sim$30 mK, while the smoothed data have an RMS of $\sim$15 mK.

\begin{figure}
\plotone{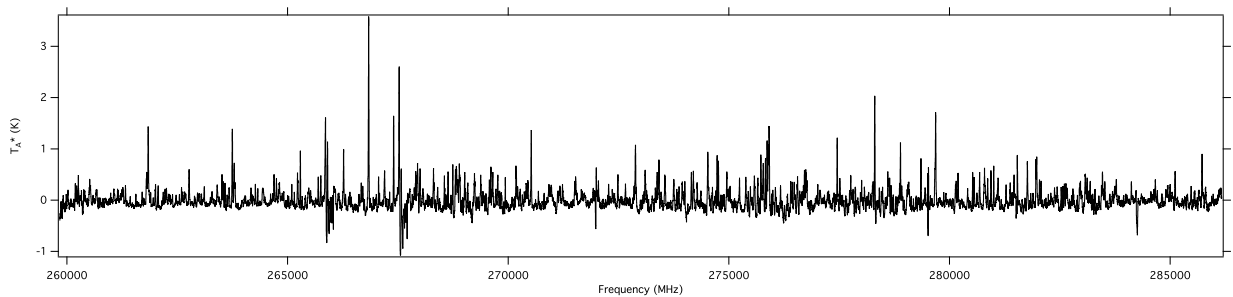}
\caption{Deconvolved and baseline-subtracted spectrum of Sgr B2(N) from 260 - 286 GHz.}
\label{full}
\end{figure}

\section{Discussion}

The preliminary reduction presented here is intended to serve as a ``quick look" at the data.  While it is likely a fairly faithful representation of the SSB spectrum, the result of the deconvolution is not unique and interested users should exercise appropriate caution in its use.  Here, we will discuss the limitations of this preliminary reduction and provide strategies for mitigating them.  These should also serve as a guide for those that wish to re-reduce the data themselves.

As discussed earlier, the robustness of the deconvolution relies heavily on having appropriate frequency sampling in the rest- and image-frequency domains for each observed frequency.  While each frequency in the bulk our observations have at least five settings, which should provide sufficient sampling, the edges of the observations do not.  In particular, care should be taken when analyzing the first and last 1 GHz of data (i.e. 260 - 261 GHz and 285 - 286 GHz).  While signals appearing in these regions are real, their assignment to the appropriate sideband is questionable.  To mitigate this, transitions from a molecule of interest should be identified in neighboring spectral windows or, in the case of the 260 - 261 GHz data, in the Nummelin et al. (1998) survey data.

The use of a third-order polynomial fit to the data to remove baselines is certainly not ideal.  While simple and effective, this method suffers from sensitivity to large changes in the average intensity within a scan, for instance, due to an abnormally strong line.  Further, the average intensity will be above the actual noise floor with the result that the zero point of the intensity scale is somewhat higher than the noise floor.  A visual inspection of the data reveals this offset is likely between 0.1 and 0.2 mK and thus the absolute intensities of the lines are expected to vary by at least this much (the relative intensities of the line peak to the surrounding baseline are likely much more robust).

Very bright lines can sometimes introduce ghosting artifacts into the SSB spectrum during deconvolution.  Given the brightest observed intensity in this spectrum is only $\sim$3 K (at 266.838 GHz [likely CH$_3$OH]), these ghosts are unlikely to be of any real concern, however, a meticulous reduction would remove the strongest signals from the spectrum prior to deconvolution, then manually stitch the signals back in afterwards.  The trade-off is a slight increase in RMS noise in the image band corresponding to the removed line, as less data were available at that frequency during the deconvolution.

Finally, the resultant SSB spectrum was visually compared to the sum total of all DSB spectra.  It was during this process that several scans were identified as inconsistent.  In general, these were cases of 4 out of 5 settings at a particular frequency indicating the presence of a line in that sideband, while the fifth did not.  The most common cause of this is  bad lock in the receiver frequencies.  While this visual inspection was thorough, due to the rough nature of this reduction, the interested user should take care to check the DSB spectra themselves against any particularly weak molecular signals.  This is especially true for the identification of new molecular signals.

A more robust reduction and analysis is planned to follow in the coming months. Until then, interested users are encouraged to use (and re-reduce) the existing dataset as they see fit.  We ask only that you cite our work as the data source and please notify us by e-mail of any resulting publications or presentations. Those wishing for further observational or data-related details should contact the authors directly.

\acknowledgments

The authors are grateful to S. Radford and the CSO staff for their assistance and the allocation of time on the telescope.  B.A.M. gratefully acknowledges funding by an NSF Graduate Research Fellowship.  The National Radio Astronomy Observatory is a facility of the National Science Foundation operated under cooperative agreement by Associated Universities, Inc.

\begin{figure}
\plotone{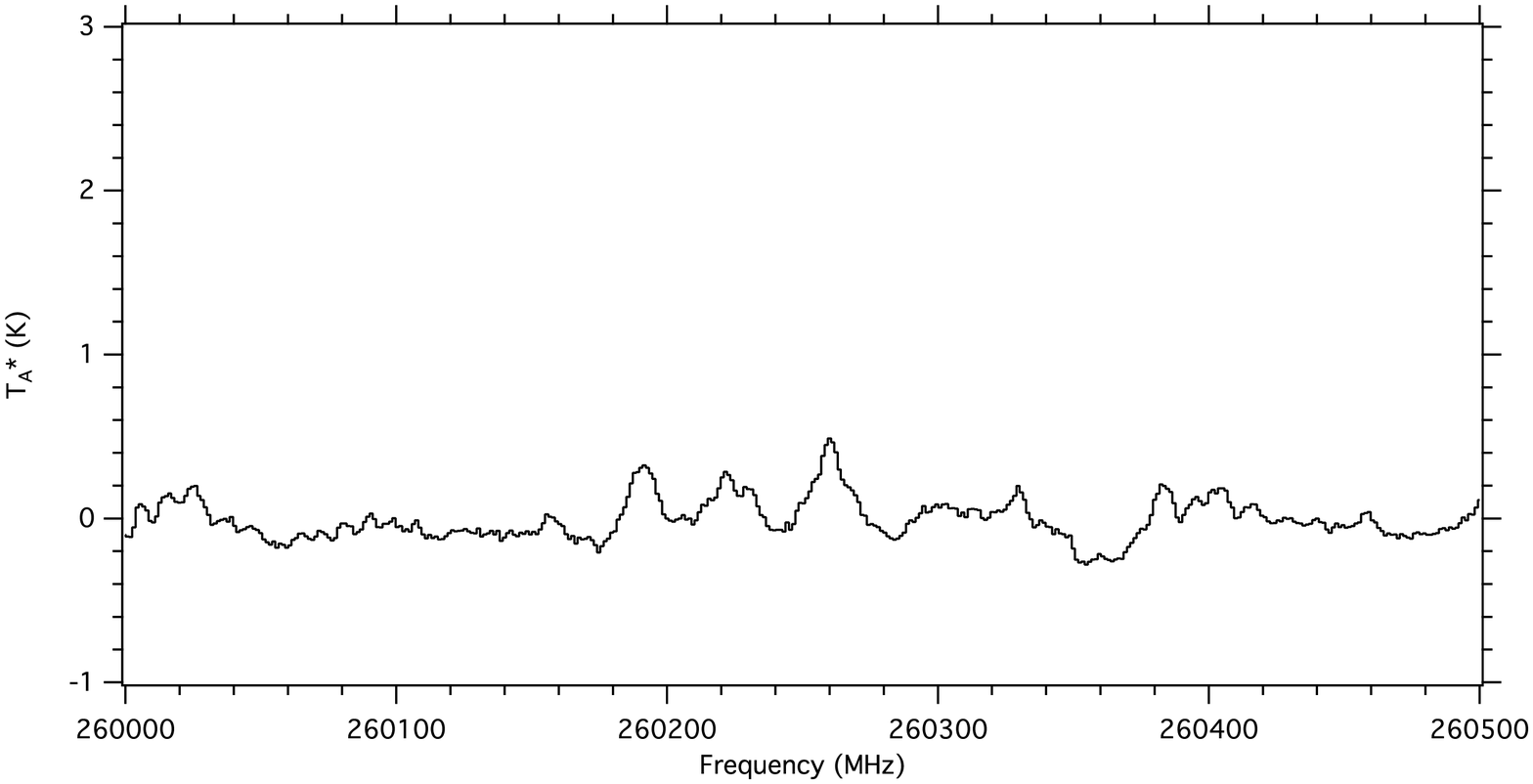}
\caption{Spectrum of Sgr B2(N) from 260.0 - 260.5 GHz}
\end{figure}

\begin{figure}
\plotone{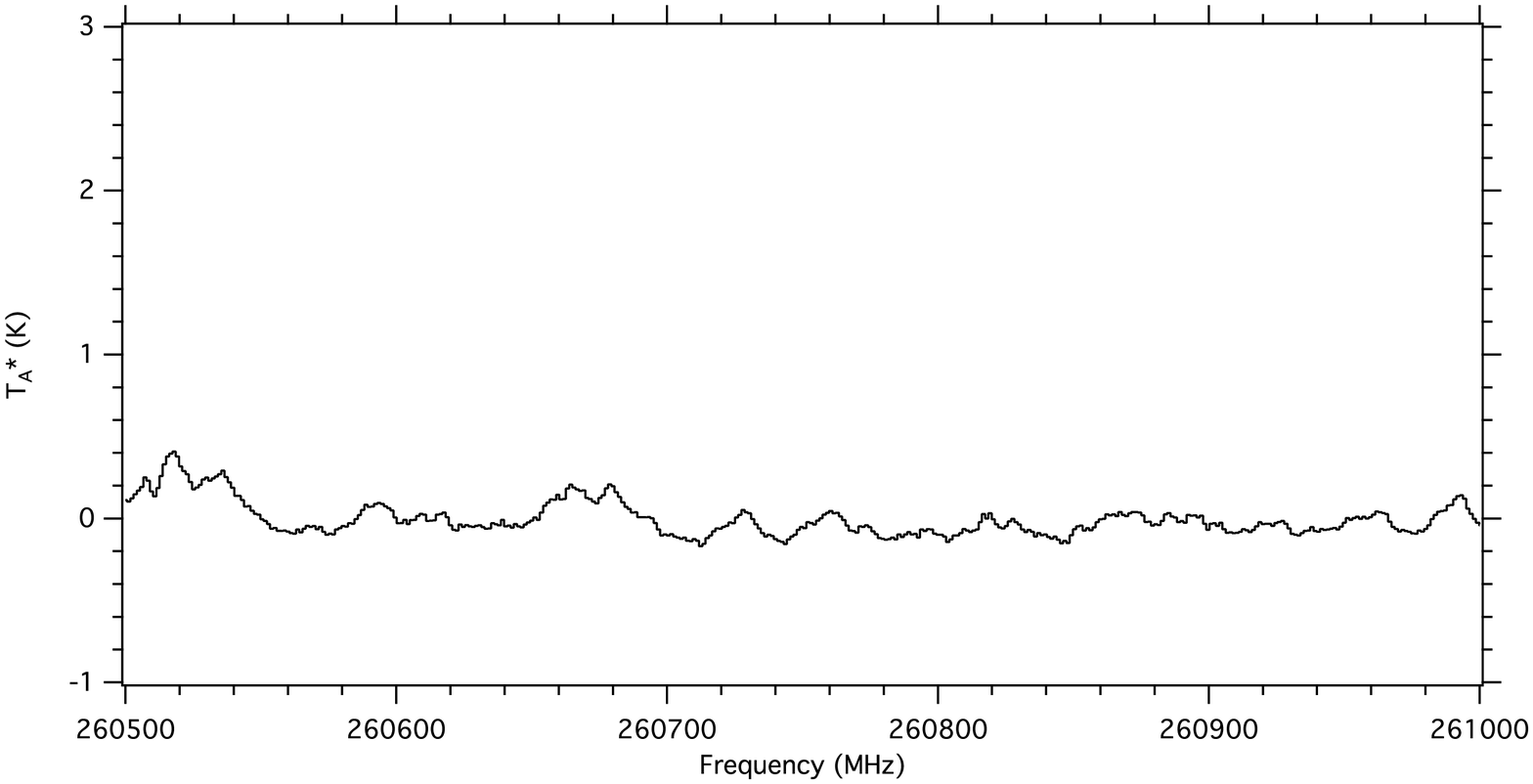}
\caption{Spectrum of Sgr B2(N) from 260.5 - 261.0 GHz}
\end{figure}

\clearpage

\begin{figure}
\plotone{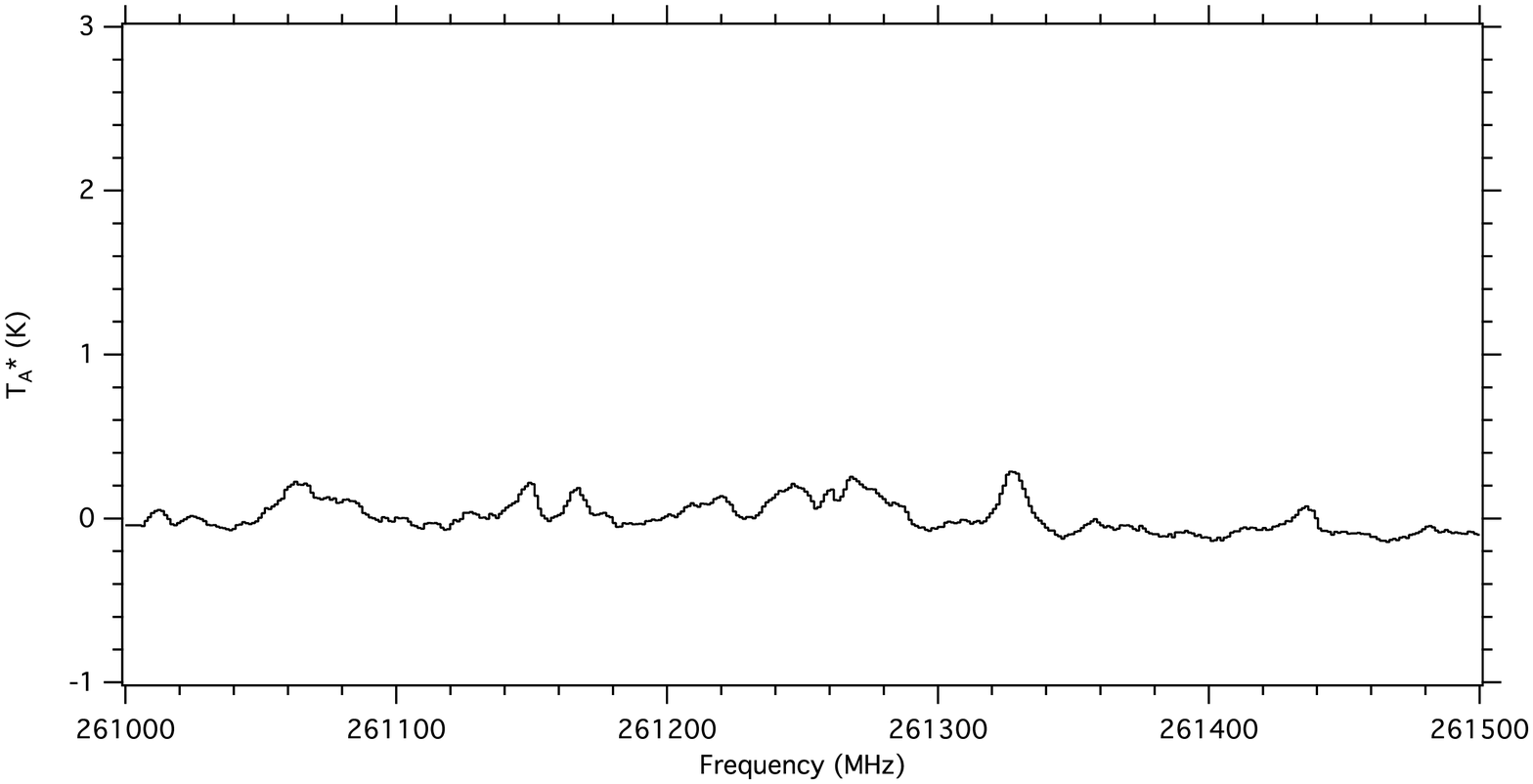}
\caption{Spectrum of Sgr B2(N) from 261.0 - 261.5 GHz}
\end{figure}

\begin{figure}
\plotone{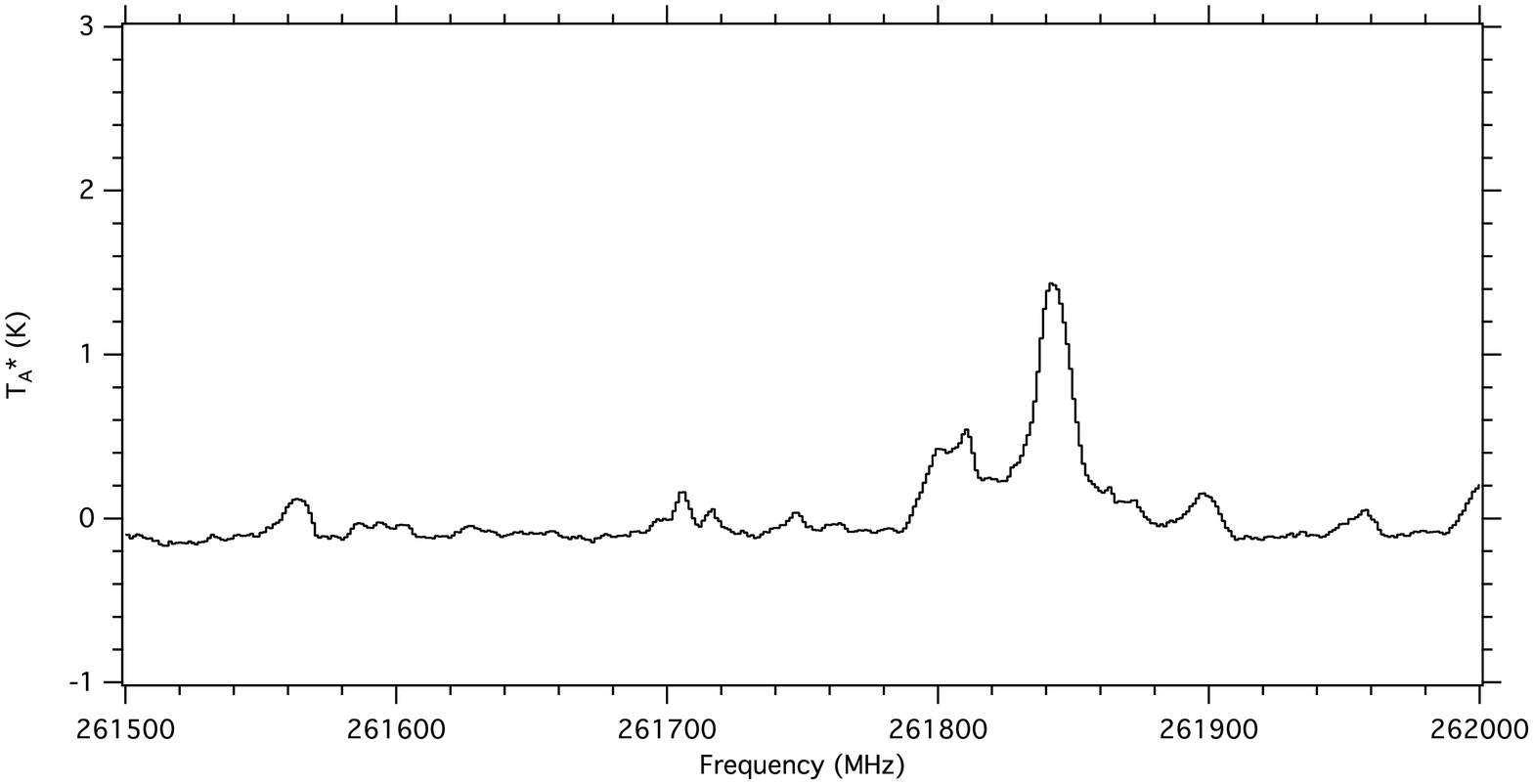}
\caption{Spectrum of Sgr B2(N) from 261.5 - 262.0 GHz}
\end{figure}

\clearpage

\begin{figure}
\plotone{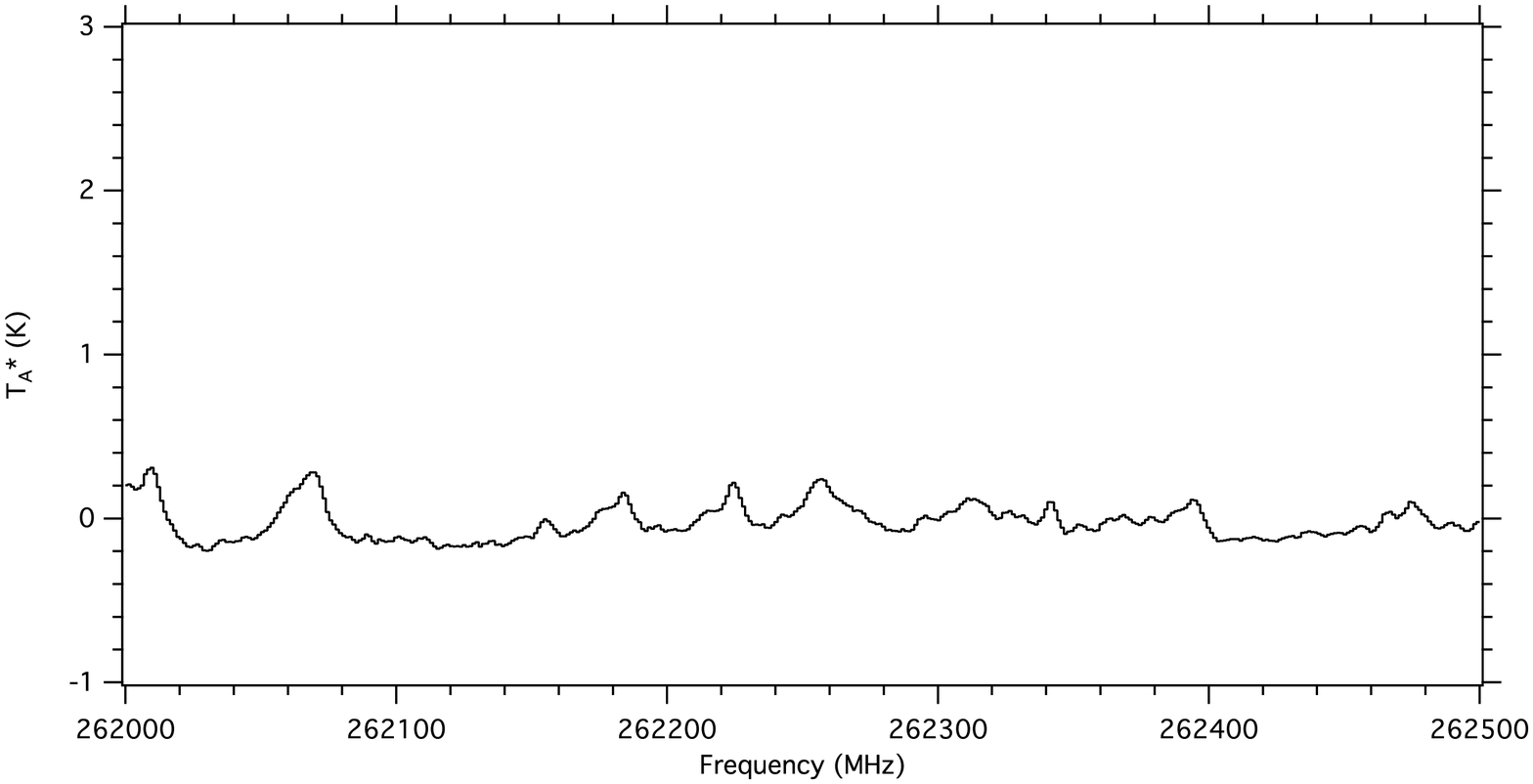}
\caption{Spectrum of Sgr B2(N) from 262.0 - 262.5 GHz}
\end{figure}

\begin{figure}
\plotone{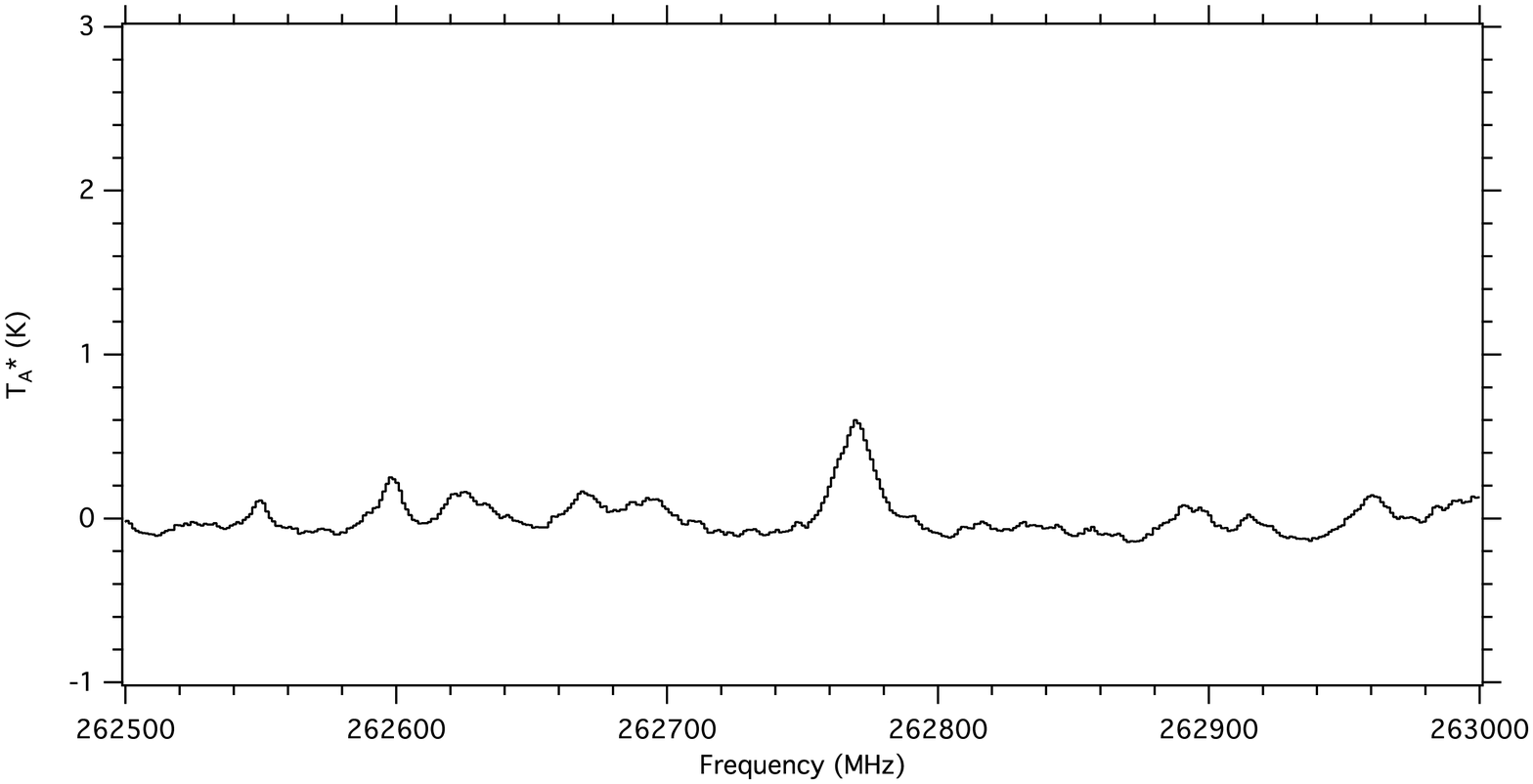}
\caption{Spectrum of Sgr B2(N) from 262.5 - 263.0 GHz}
\end{figure}

\clearpage

\begin{figure}
\plotone{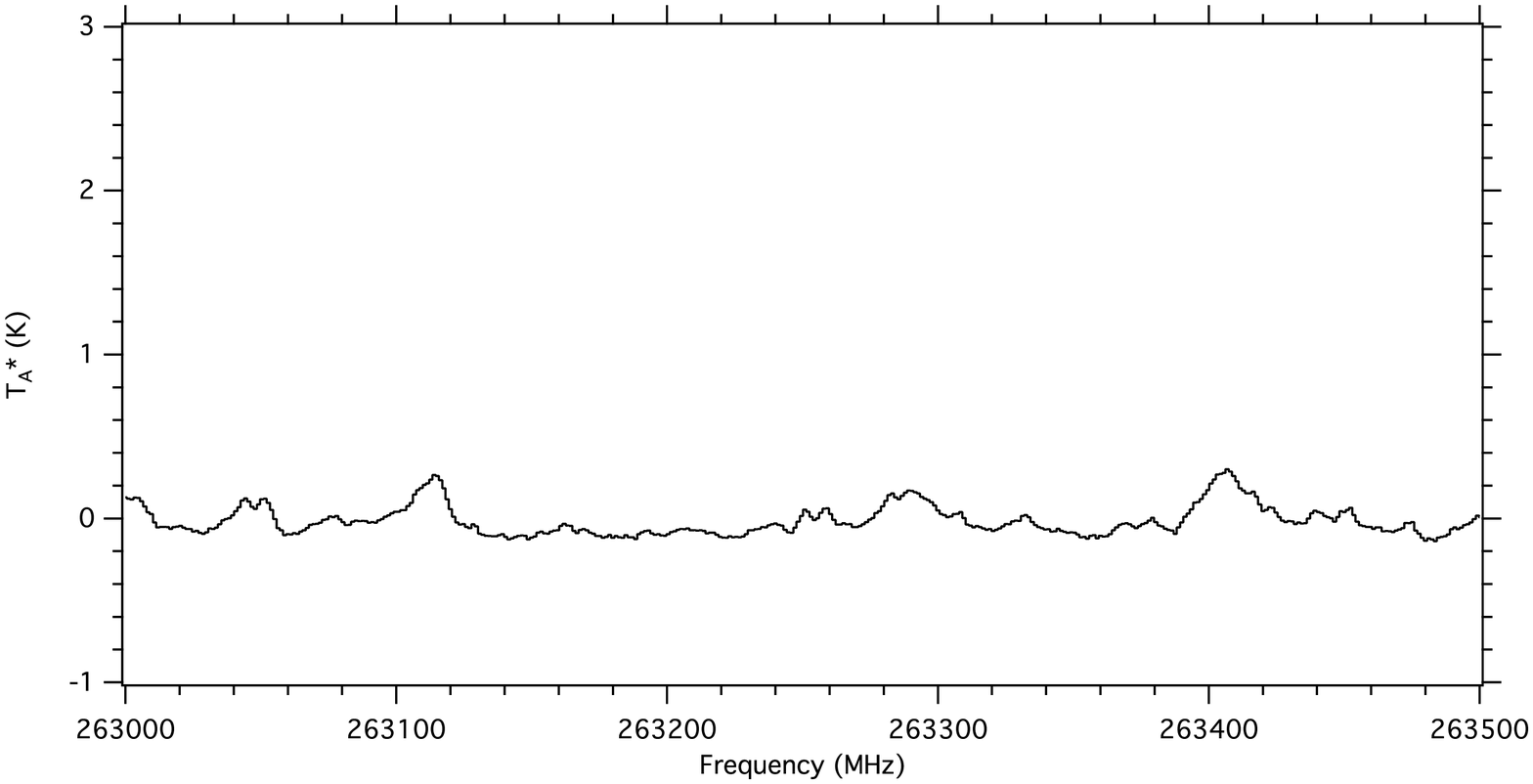}
\caption{Spectrum of Sgr B2(N) from 263.0 - 263.5 GHz}
\end{figure}

\begin{figure}
\plotone{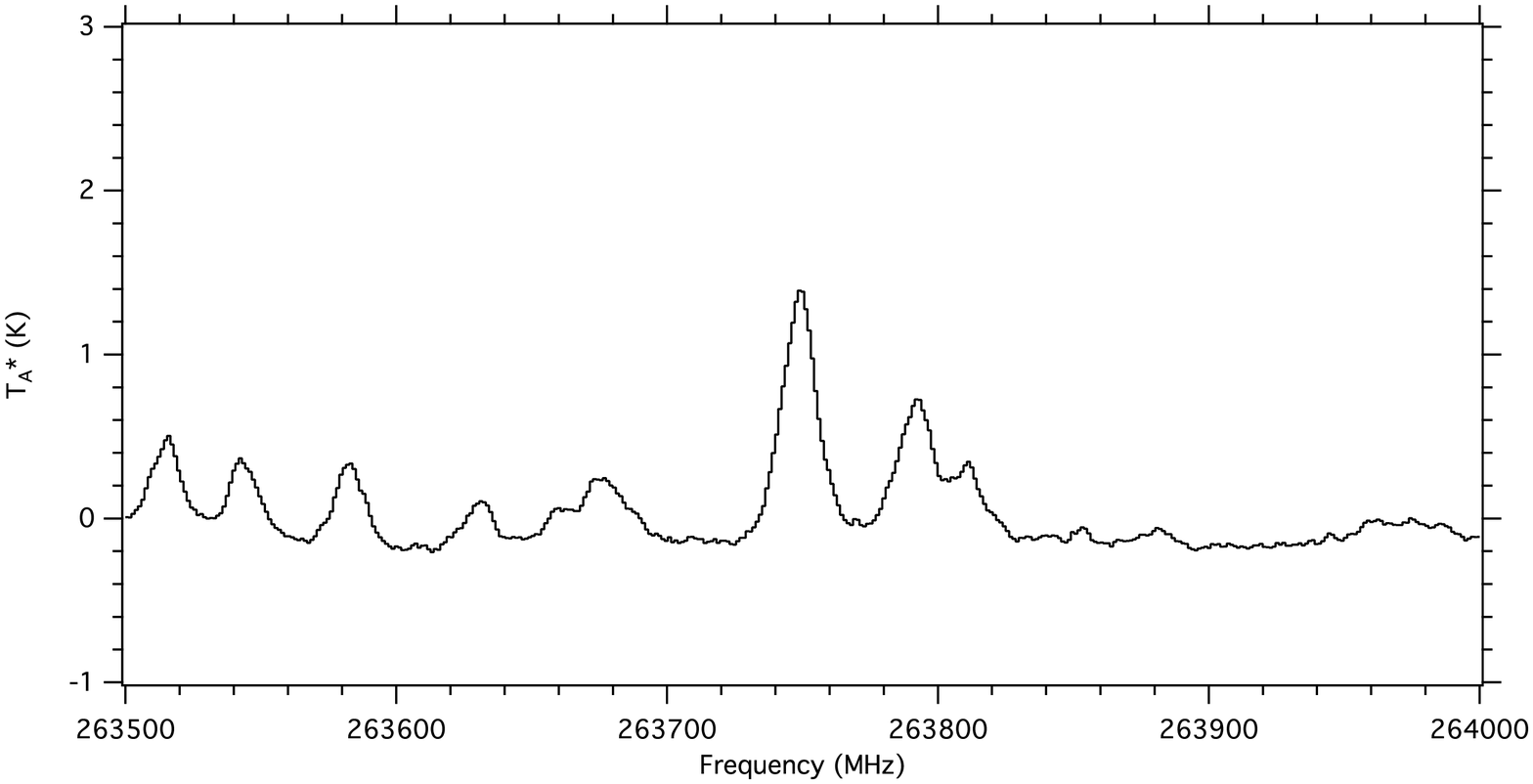}
\caption{Spectrum of Sgr B2(N) from 263.5 - 264.0 GHz}
\end{figure}

\clearpage

\begin{figure}
\plotone{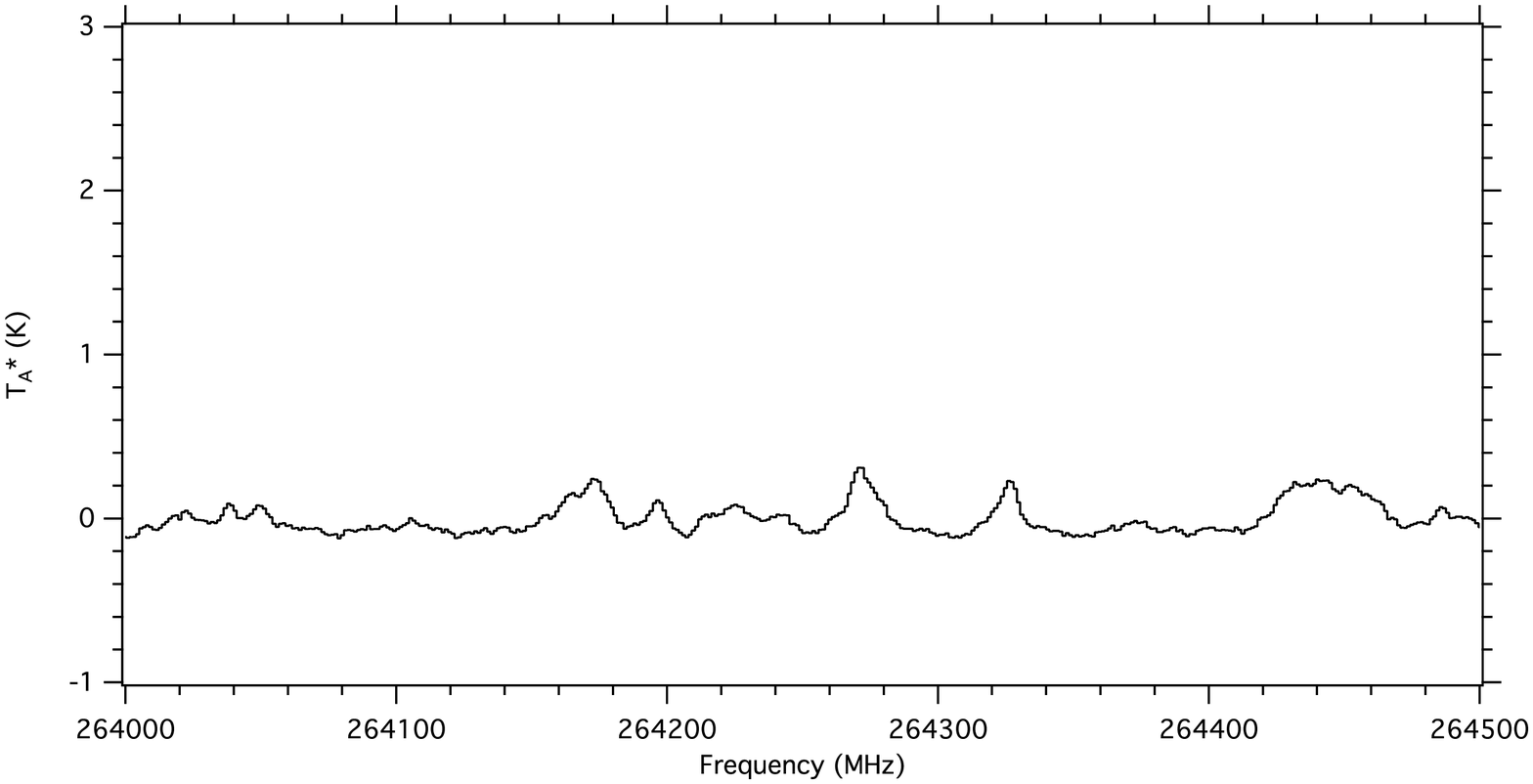}
\caption{Spectrum of Sgr B2(N) from 264.0 - 264.5 GHz}
\end{figure}

\begin{figure}
\plotone{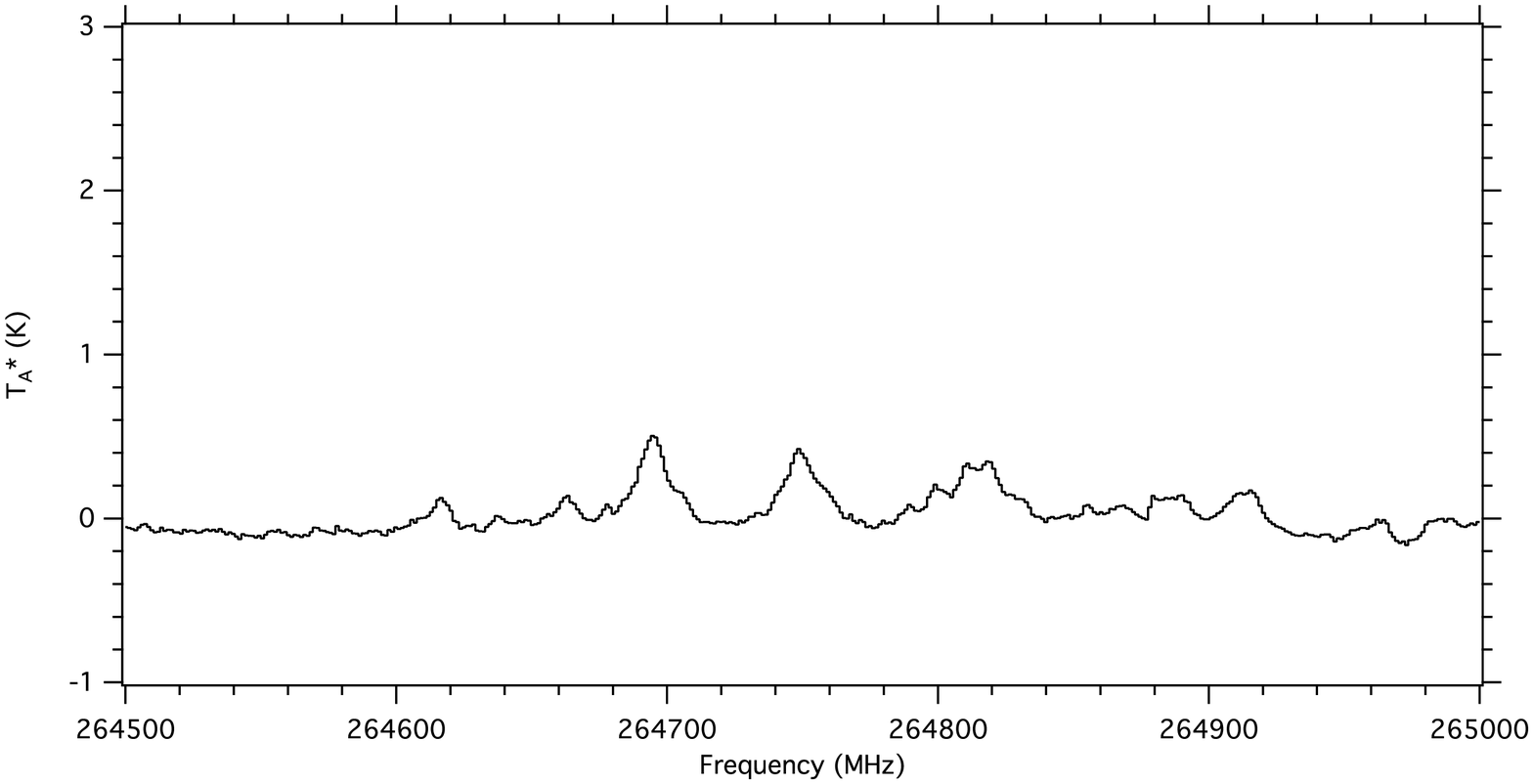}
\caption{Spectrum of Sgr B2(N) from 264.5 - 265.0 GHz}
\end{figure}

\clearpage

\begin{figure}
\plotone{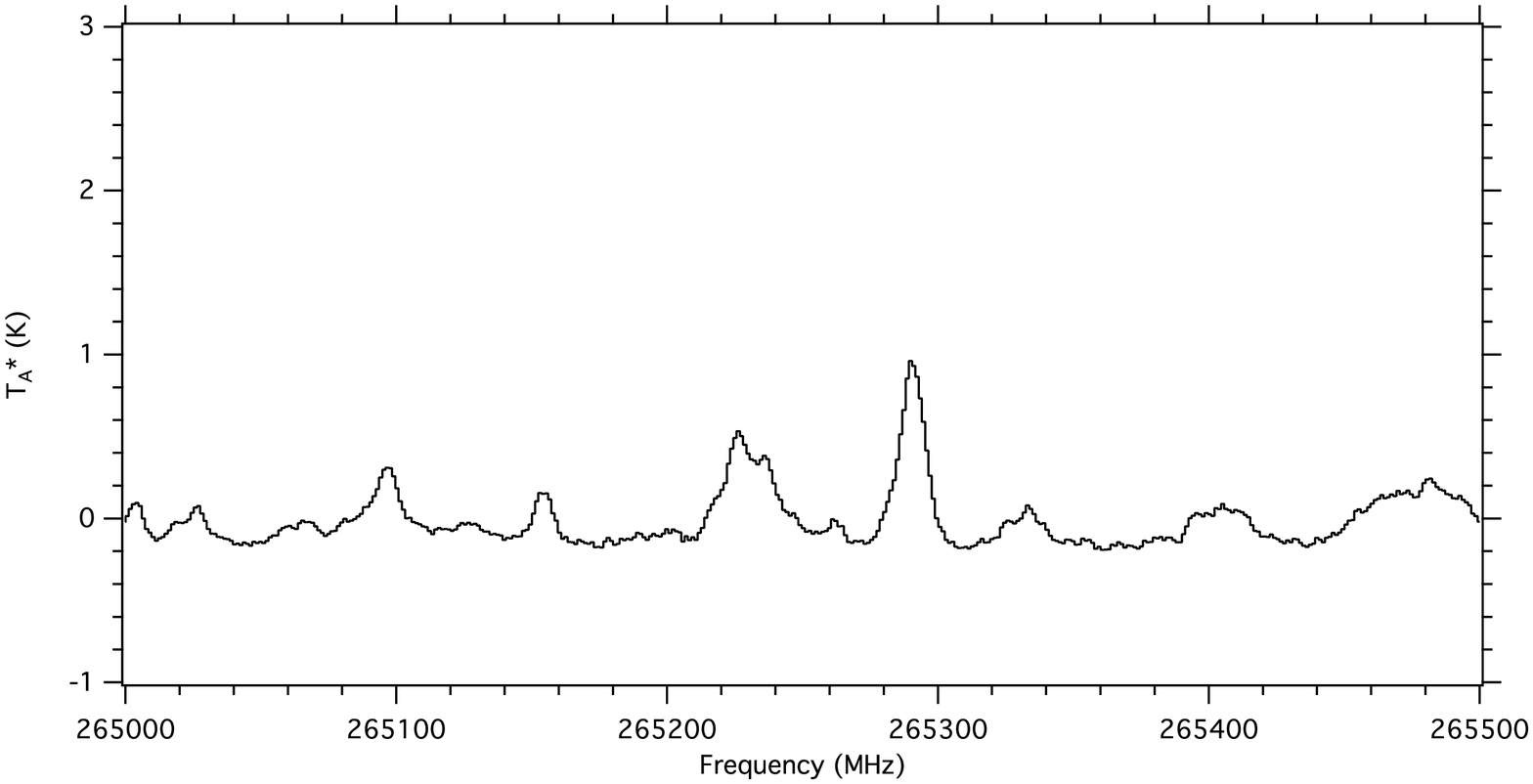}
\caption{Spectrum of Sgr B2(N) from 265.0 - 265.5 GHz}
\end{figure}

\begin{figure}
\plotone{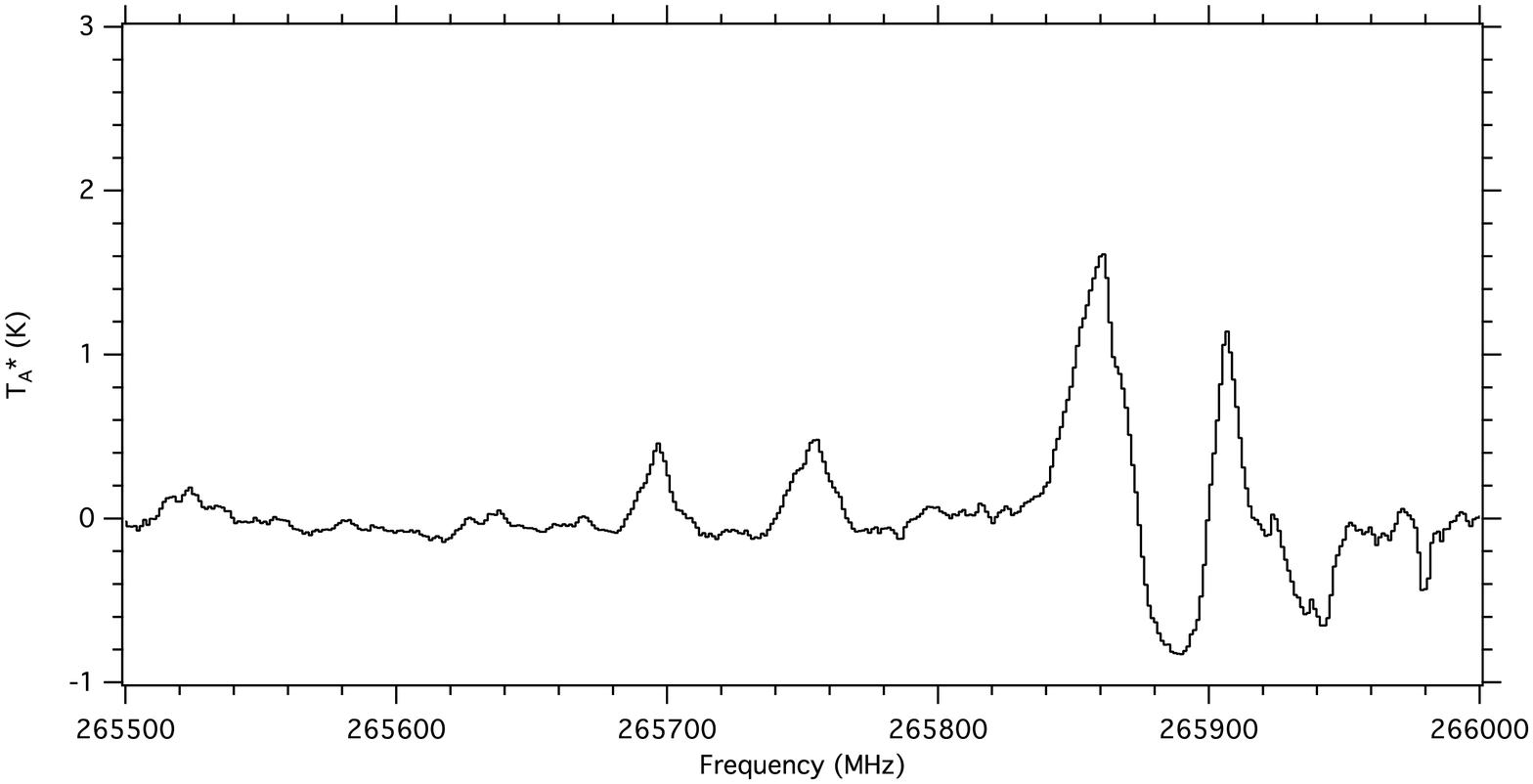}
\caption{Spectrum of Sgr B2(N) from 265.5 - 266.0 GHz}
\end{figure}

\clearpage

\begin{figure}
\plotone{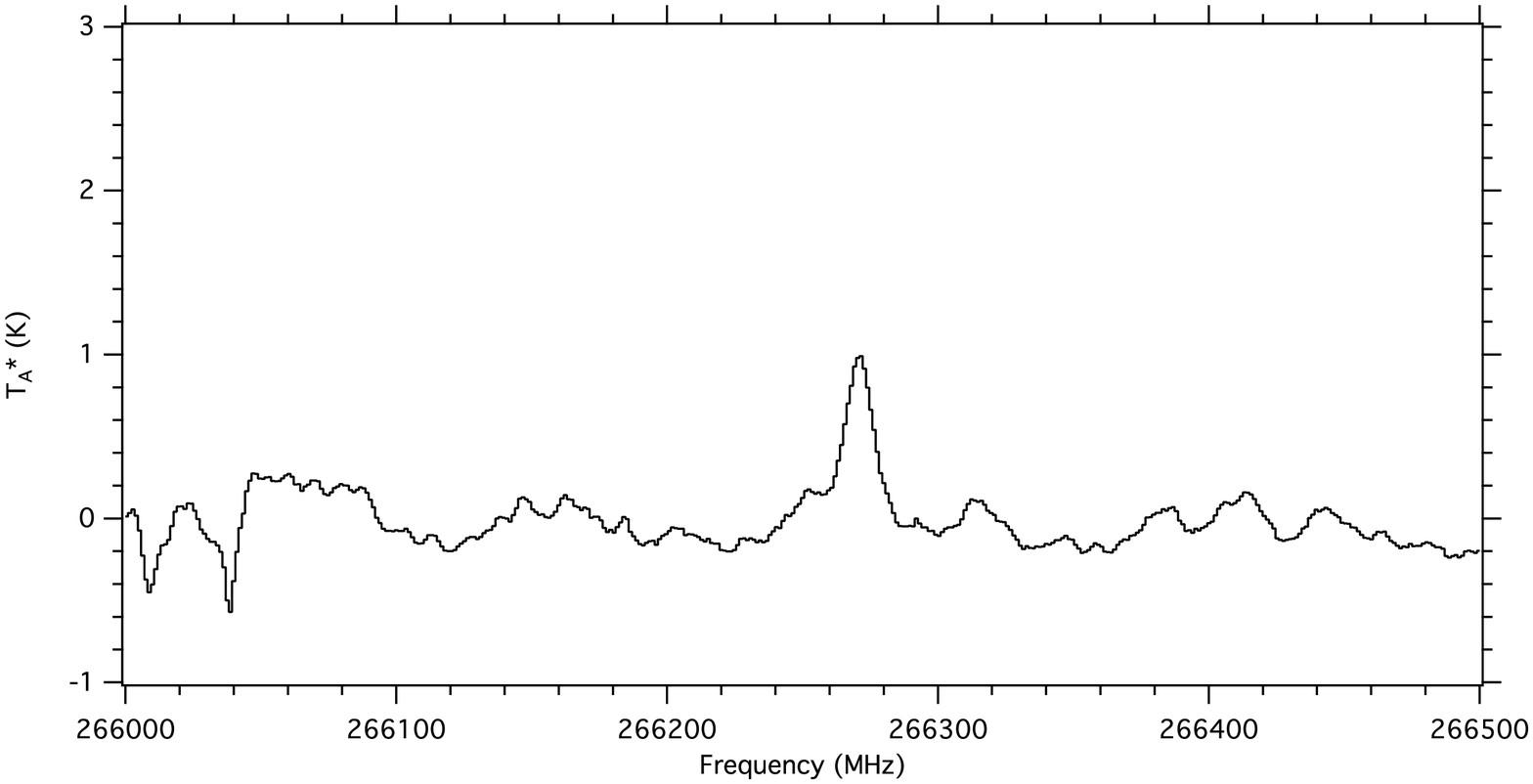}
\caption{Spectrum of Sgr B2(N) from 266.0 - 266.5 GHz}
\end{figure}

\begin{figure}
\plotone{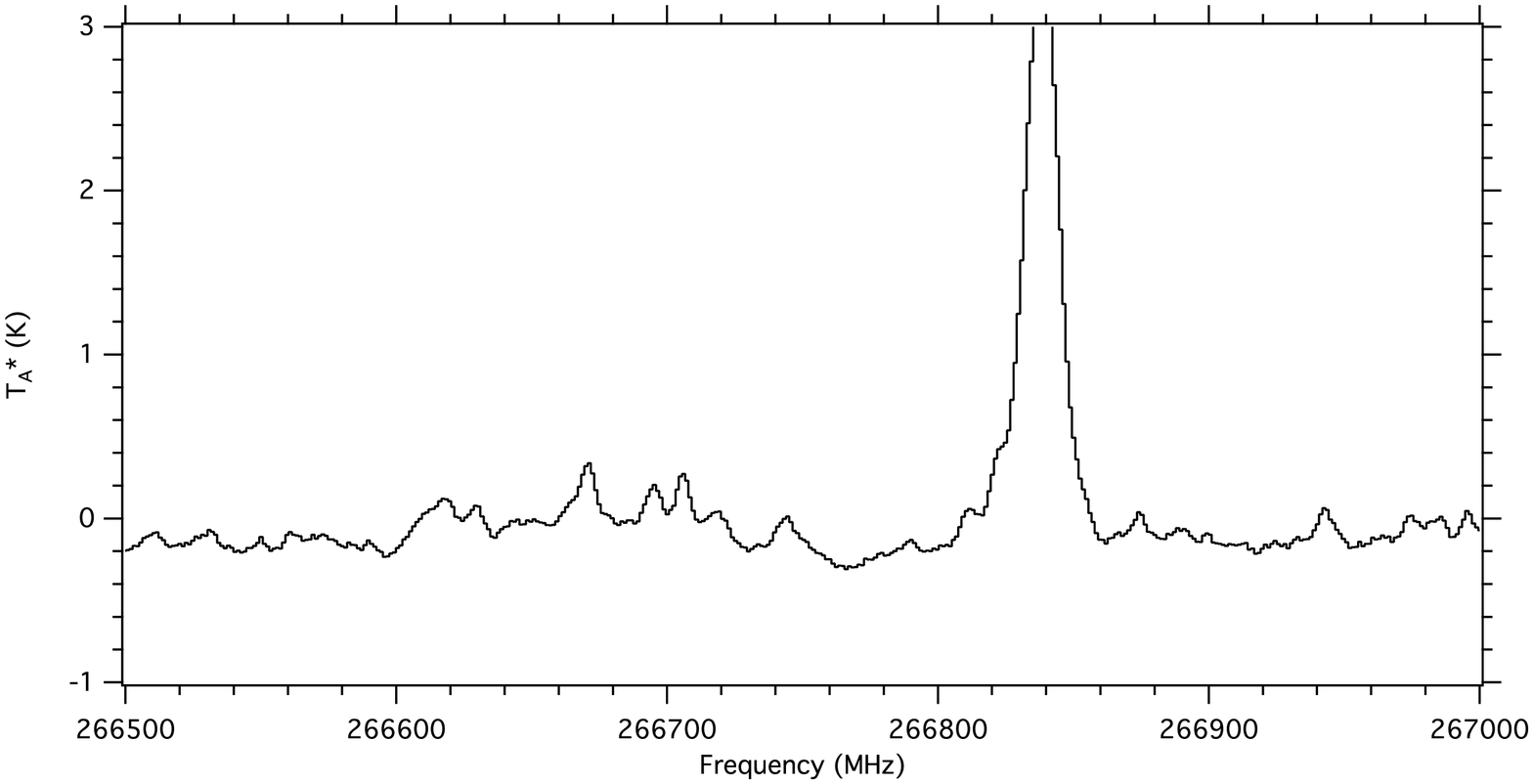}
\caption{Spectrum of Sgr B2(N) from 266.5 - 267.0 GHz}
\end{figure}

\clearpage

\begin{figure}
\plotone{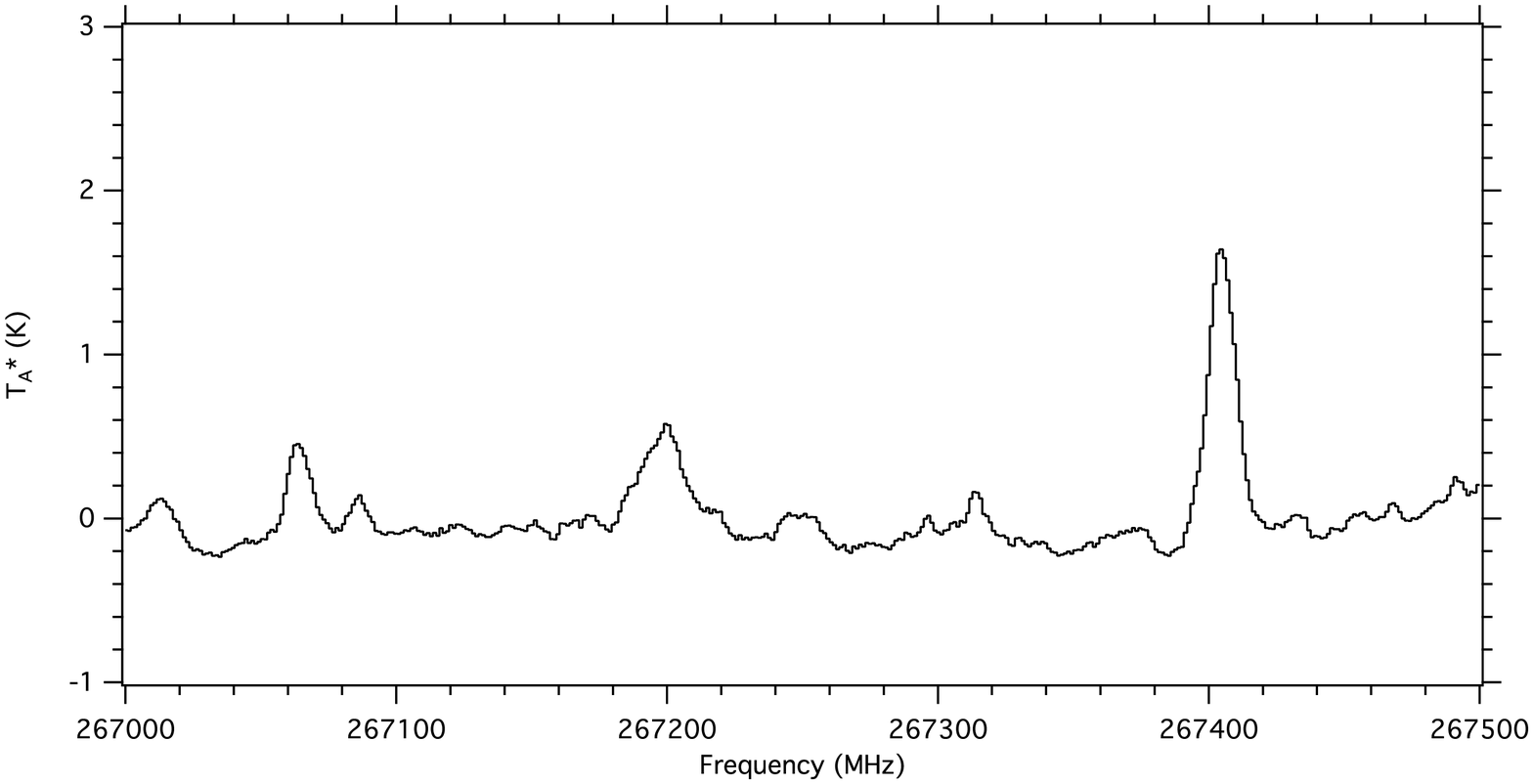}
\caption{Spectrum of Sgr B2(N) from 267.0 - 267.5 GHz}
\end{figure}

\begin{figure}
\plotone{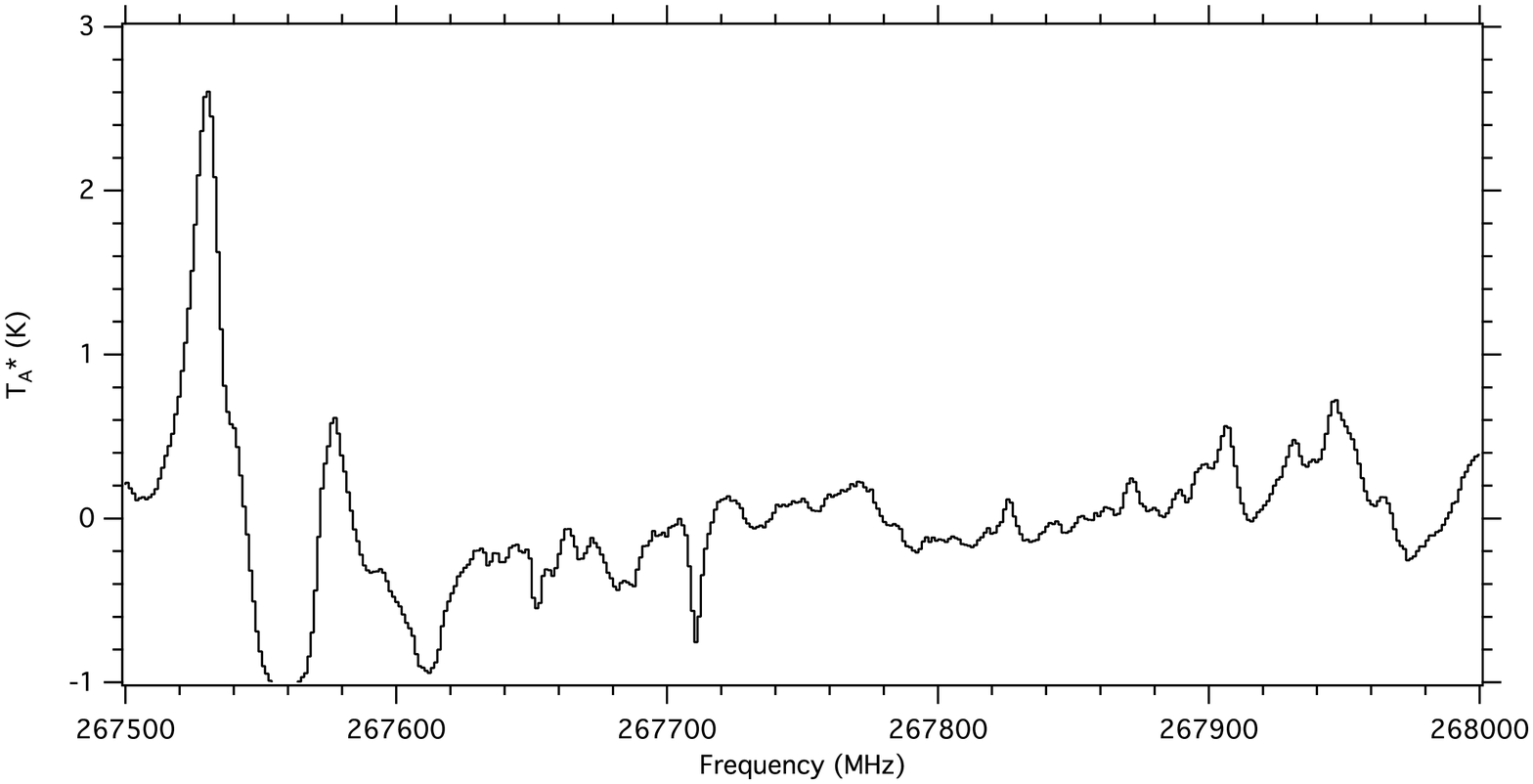}
\caption{Spectrum of Sgr B2(N) from 267.5 - 268.0 GHz}
\end{figure}

\clearpage

\begin{figure}
\plotone{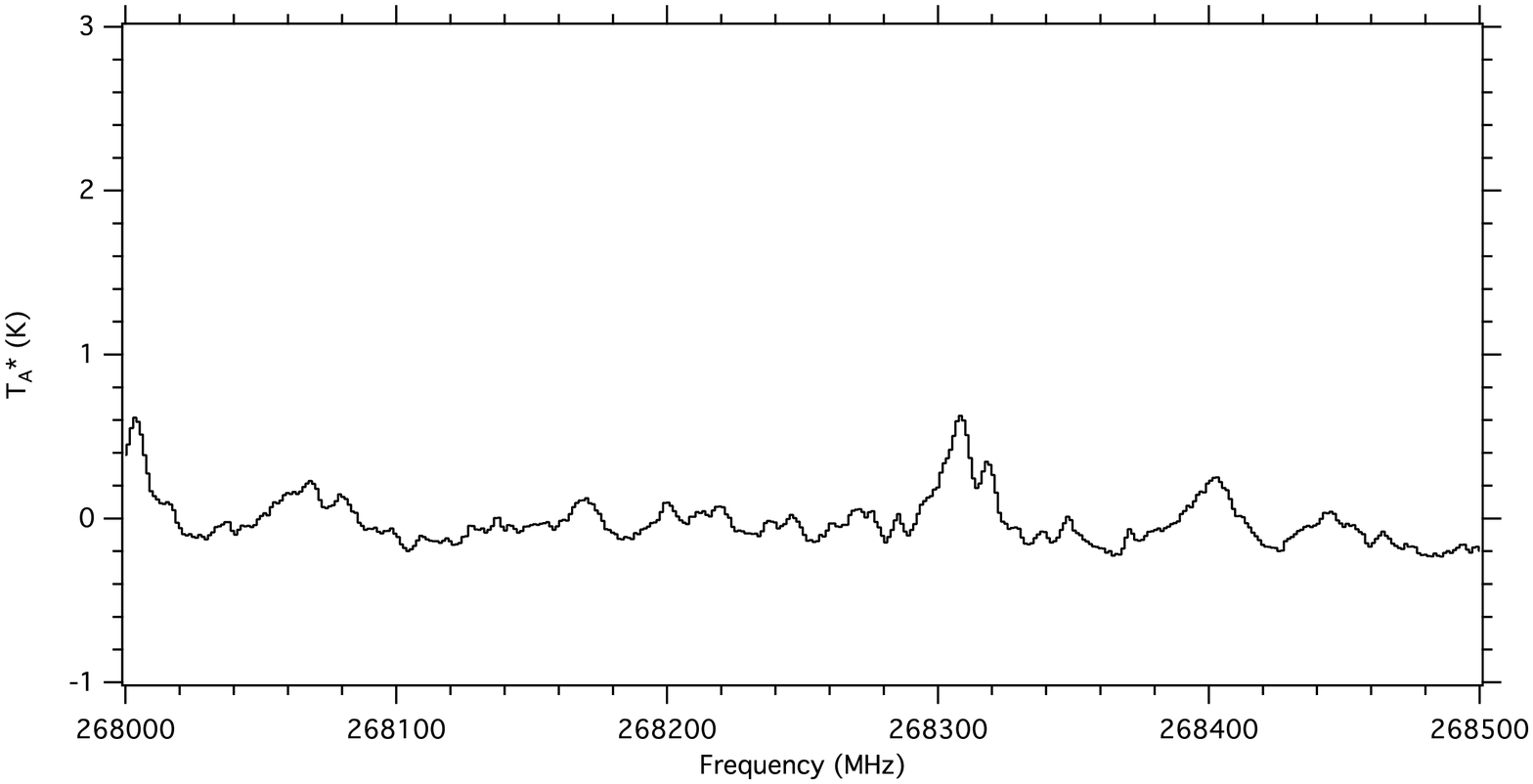}
\caption{Spectrum of Sgr B2(N) from 268.0 - 268.5 GHz}
\end{figure}

\begin{figure}
\plotone{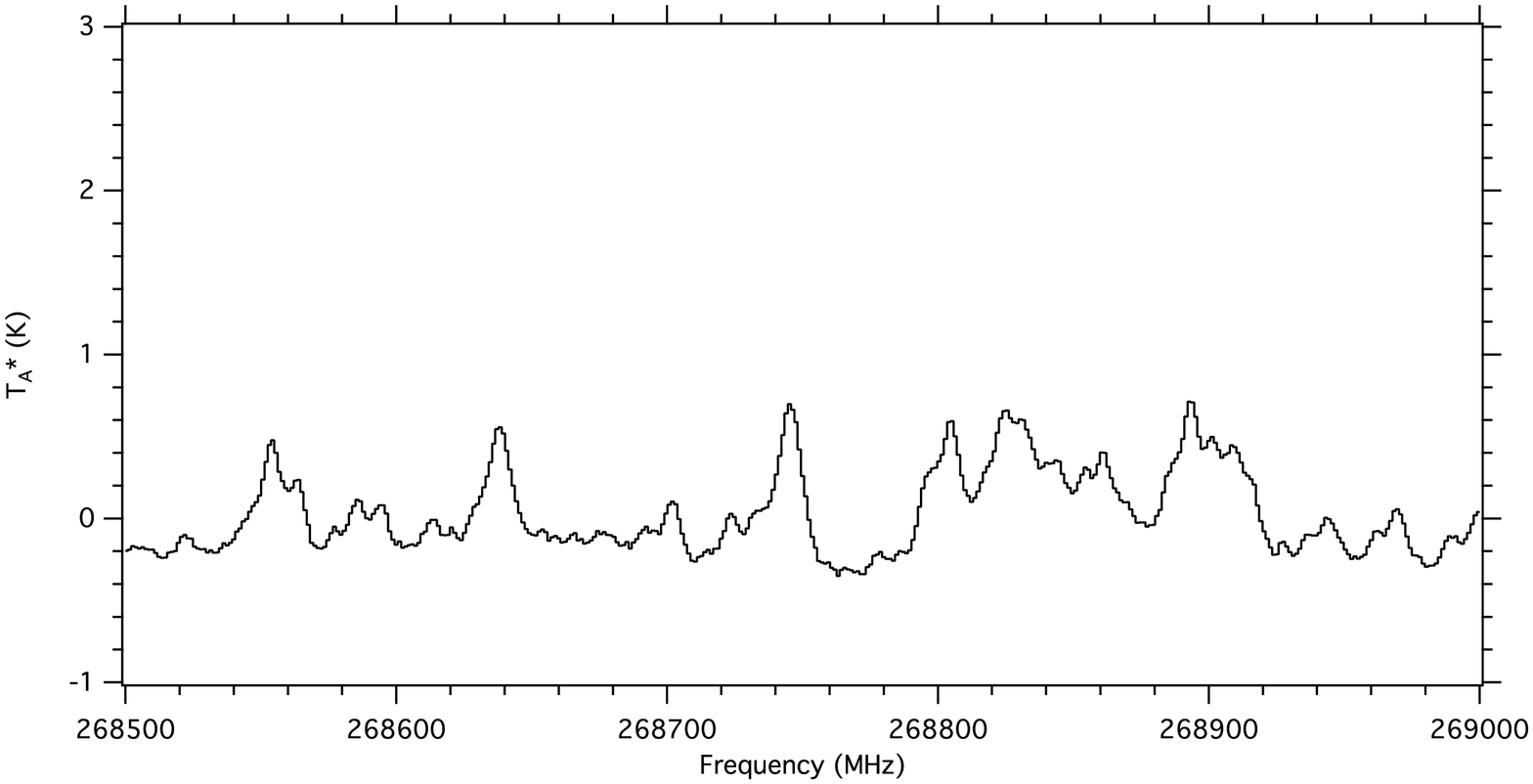}
\caption{Spectrum of Sgr B2(N) from 268.5 - 269.0 GHz}
\end{figure}

\clearpage

\begin{figure}
\plotone{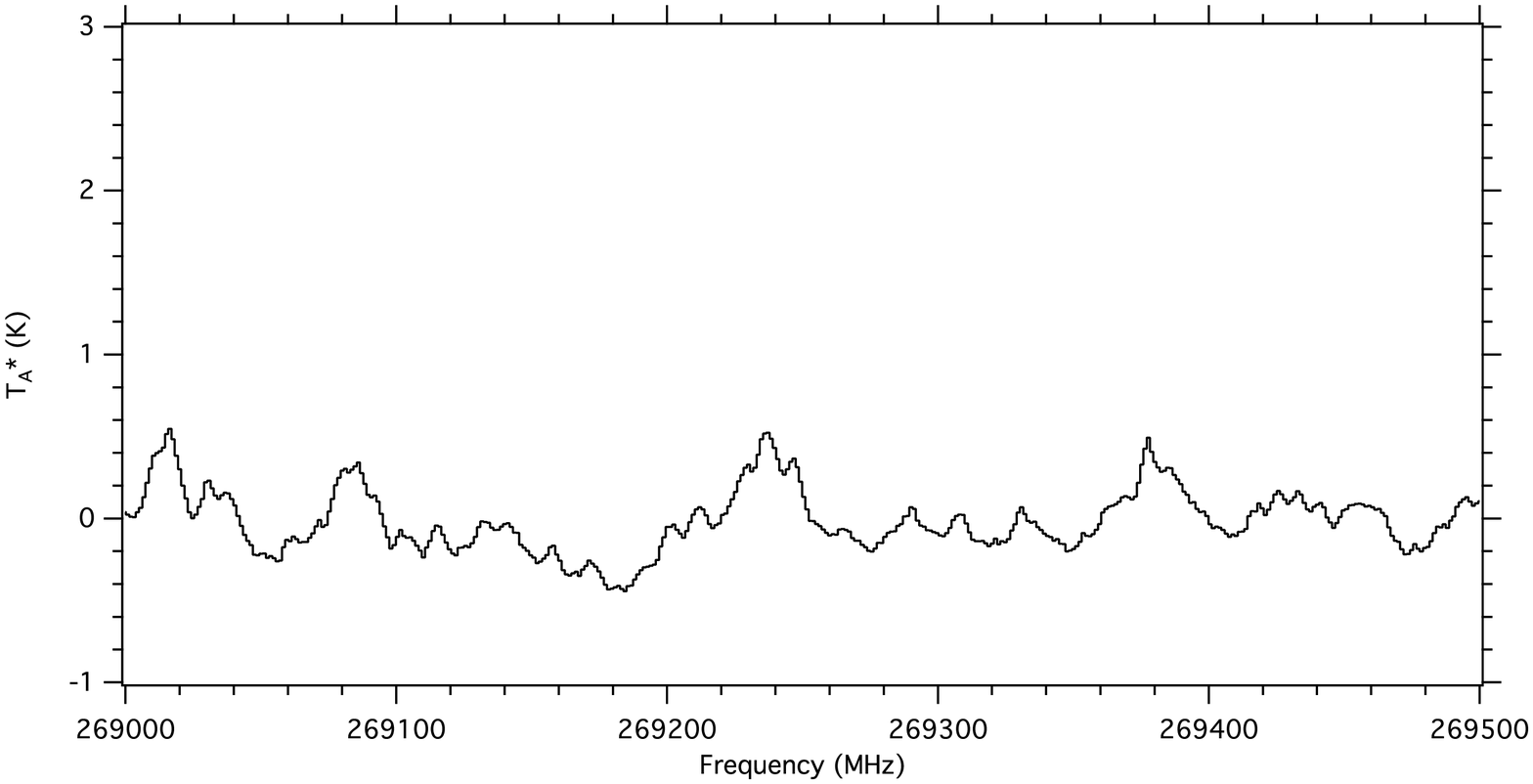}
\caption{Spectrum of Sgr B2(N) from 269.0 - 269.5 GHz}
\end{figure}

\begin{figure}
\plotone{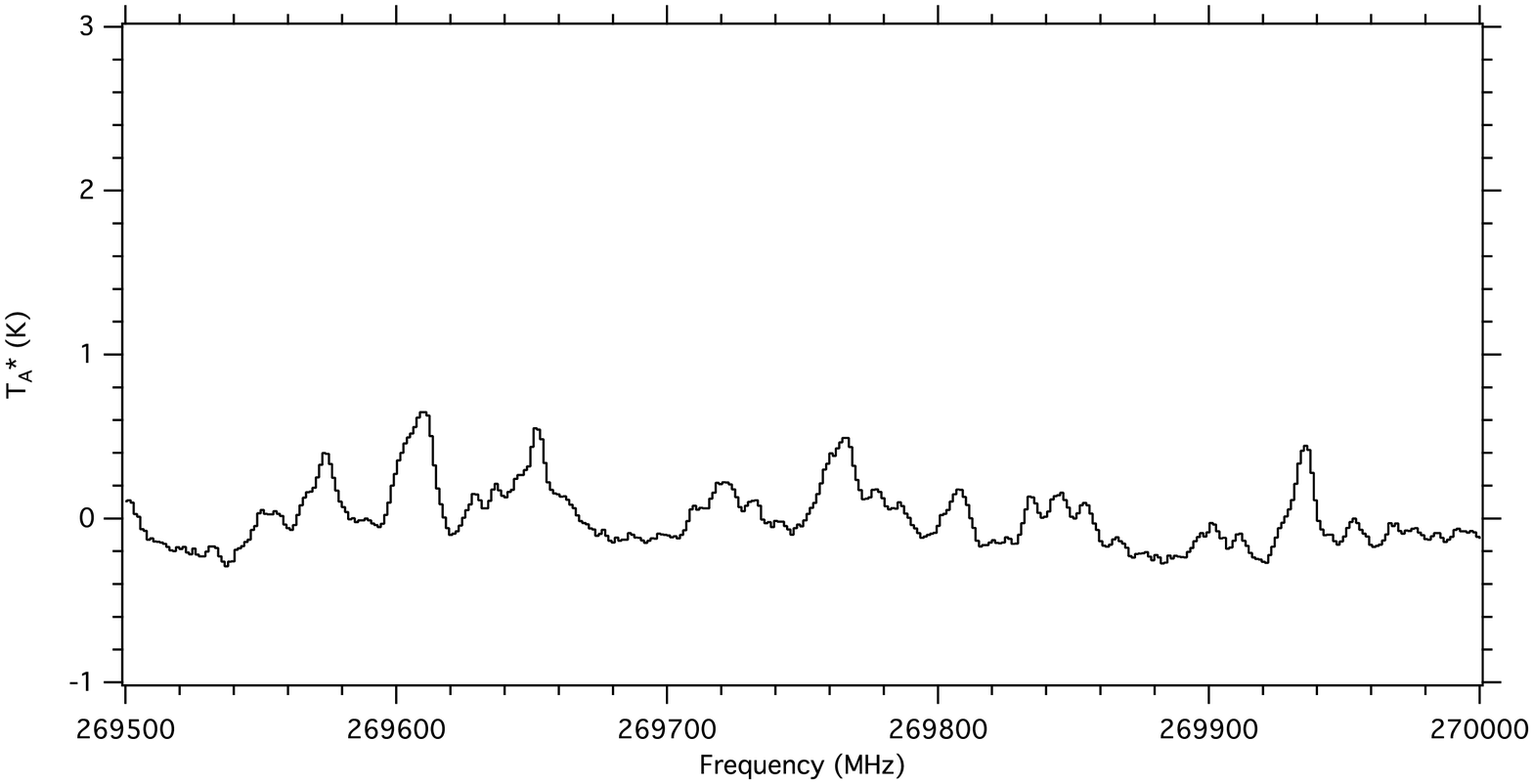}
\caption{Spectrum of Sgr B2(N) from 269.5 - 270.0 GHz}
\end{figure}

\clearpage

\begin{figure}
\plotone{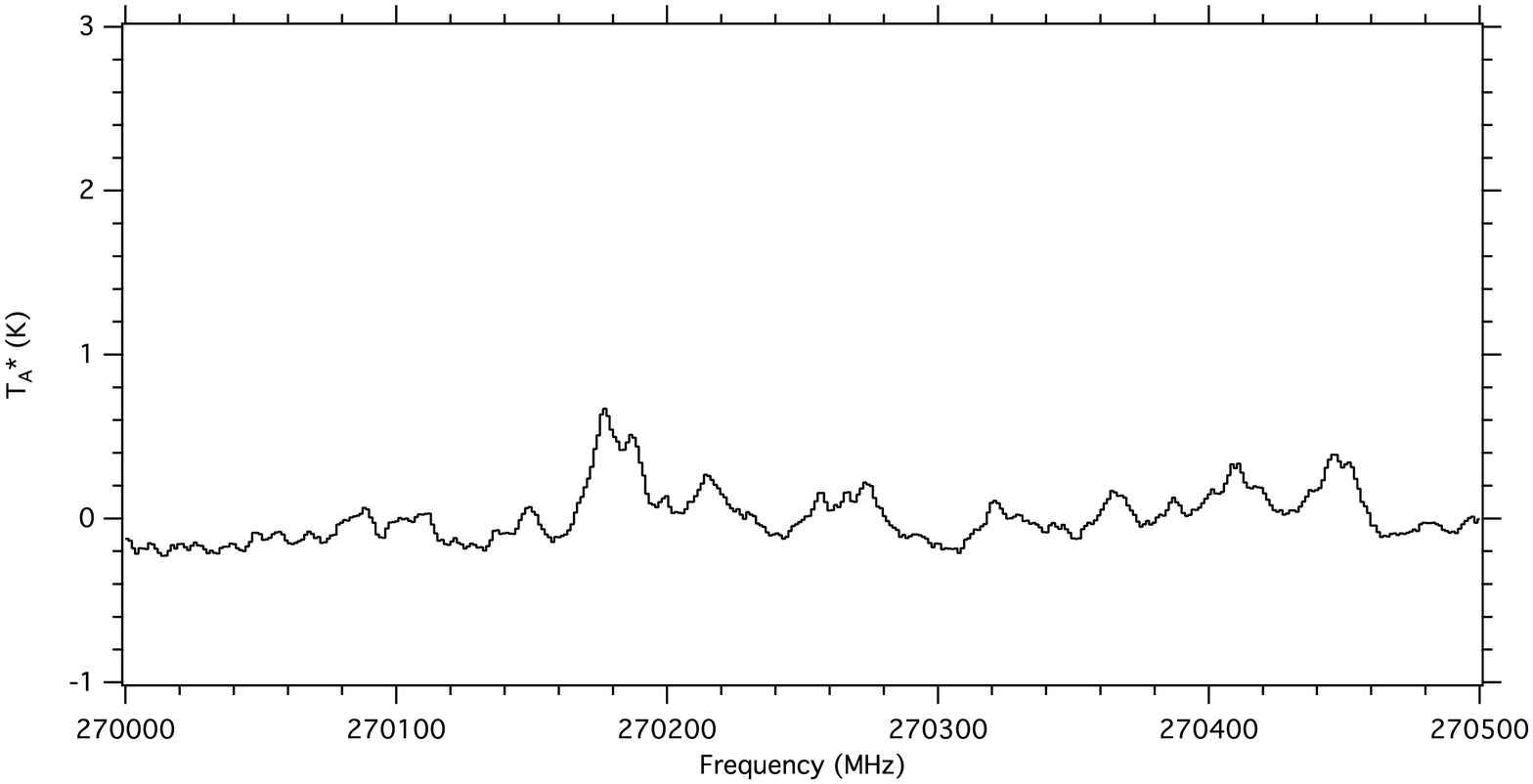}
\caption{Spectrum of Sgr B2(N) from 270.0 - 270.5 GHz}
\end{figure}

\begin{figure}
\plotone{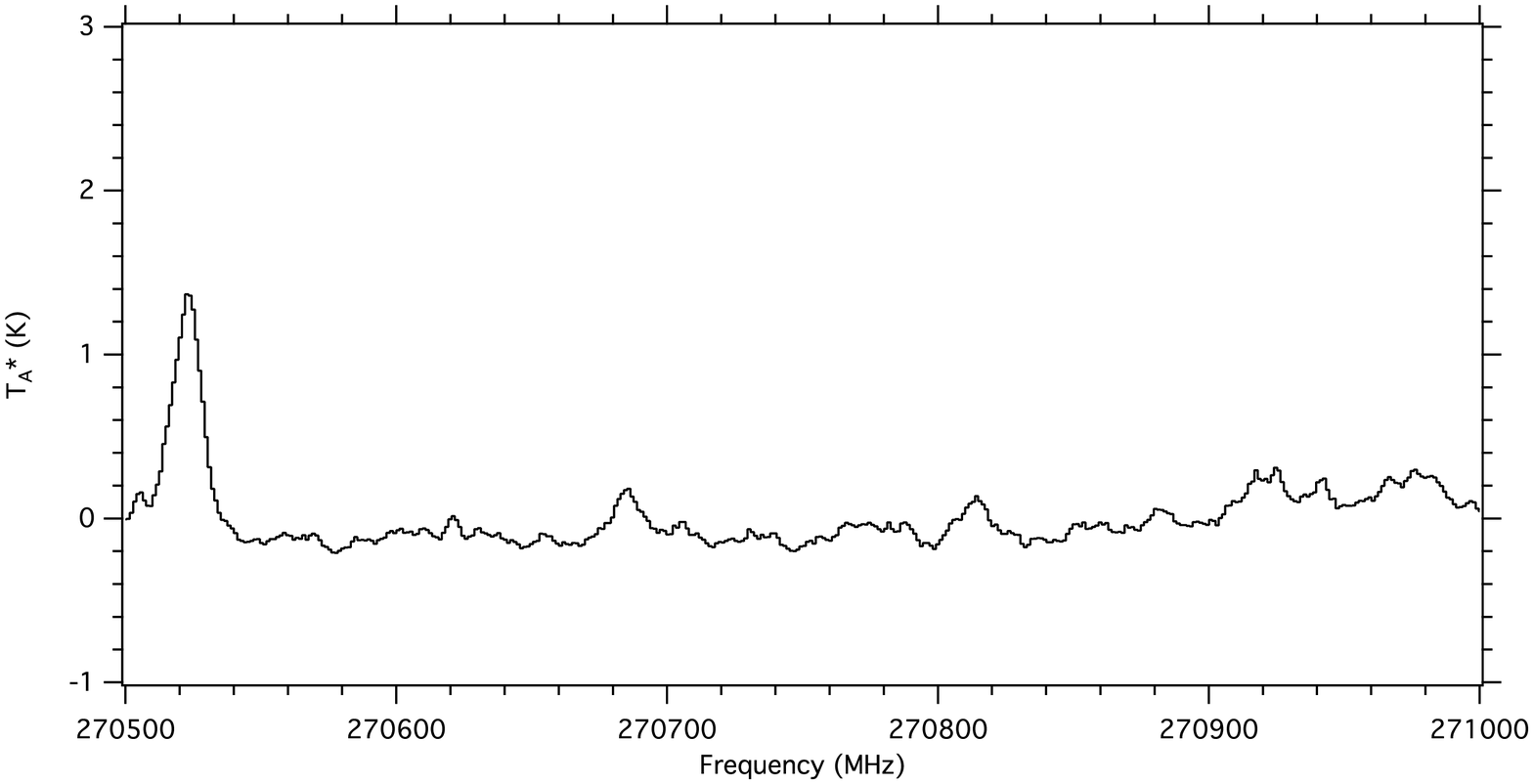}
\caption{Spectrum of Sgr B2(N) from 270.5 - 271.0 GHz}
\end{figure}

\clearpage

\begin{figure}
\plotone{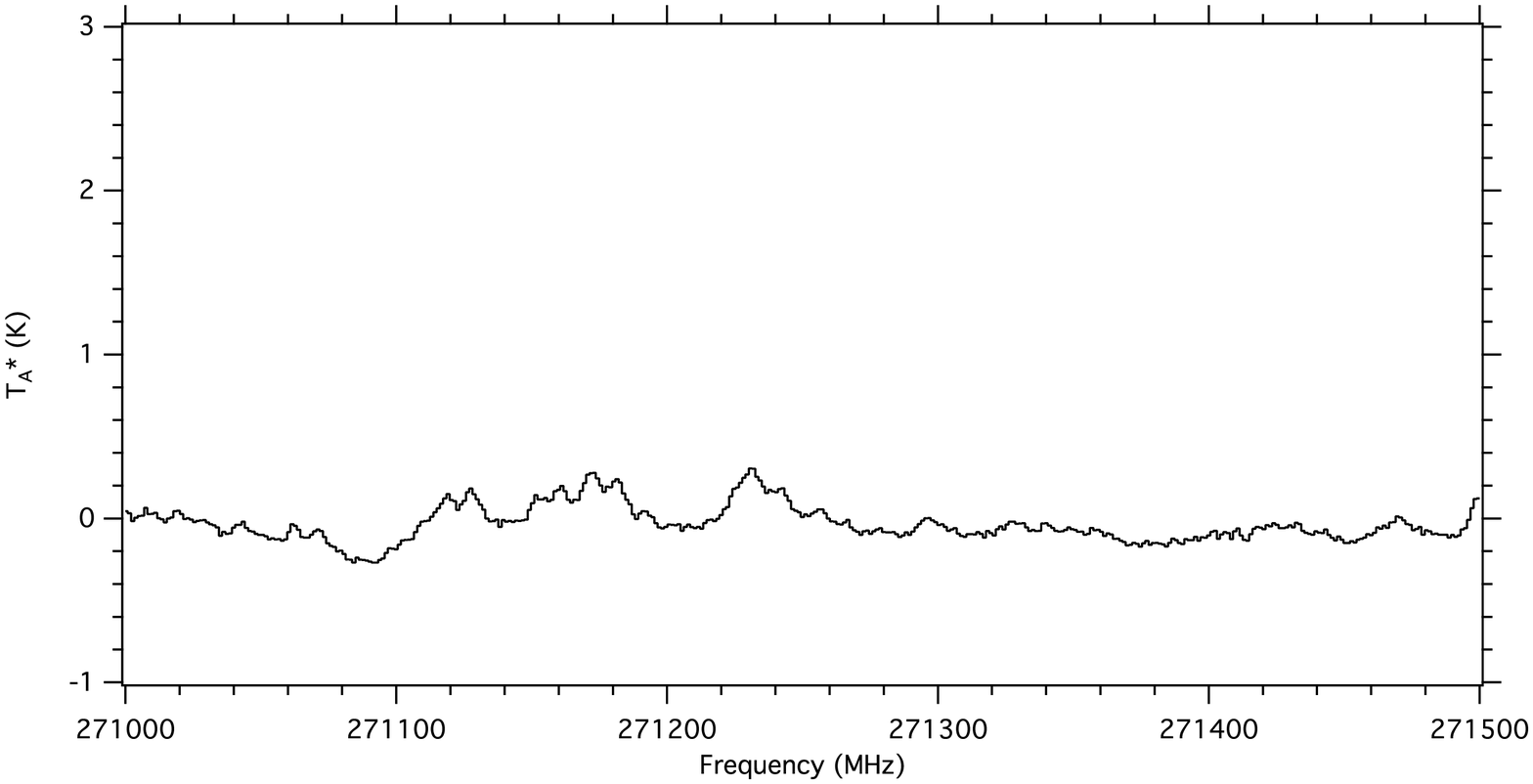}
\caption{Spectrum of Sgr B2(N) from 271.0 - 271.5 GHz}
\end{figure}

\begin{figure}
\plotone{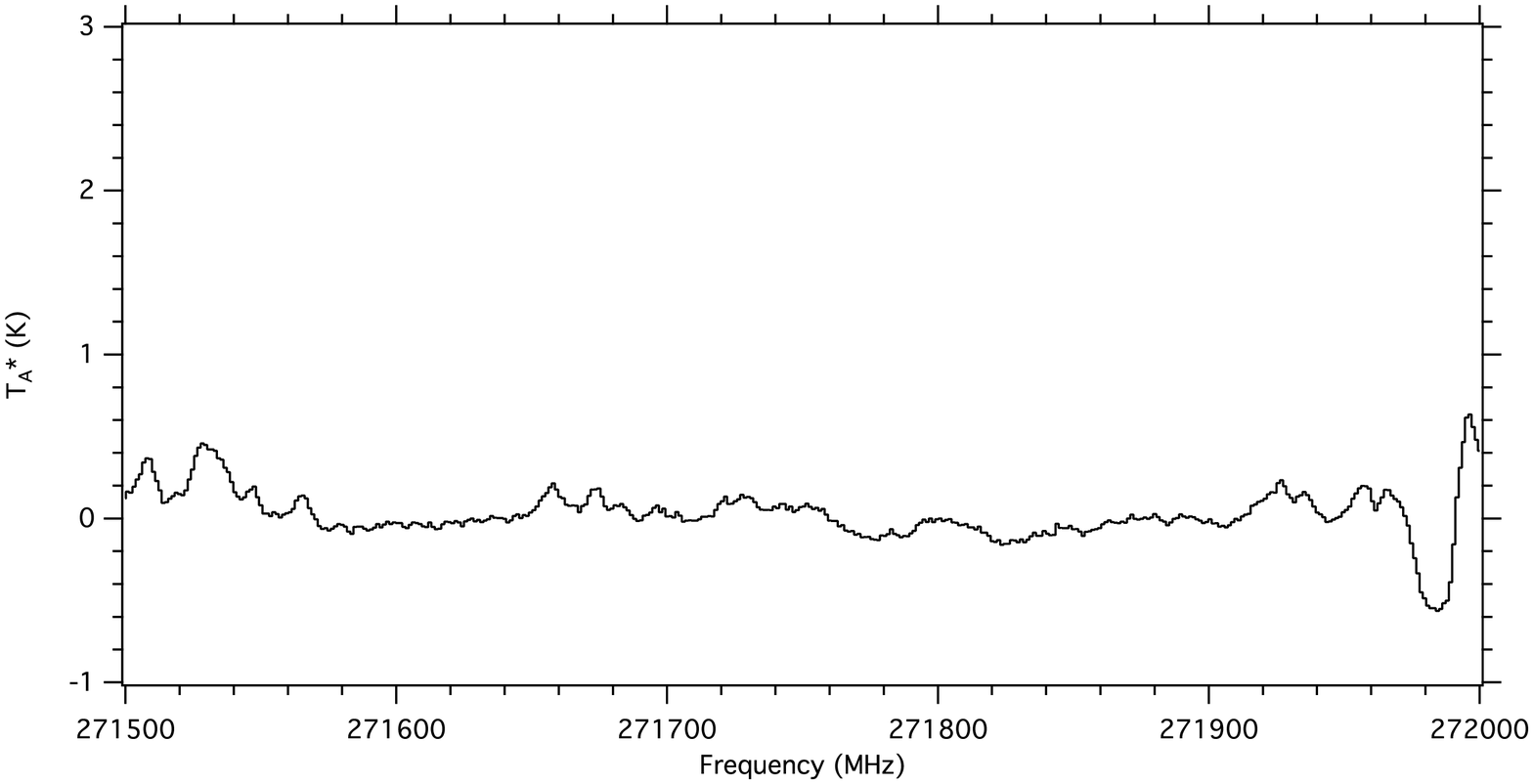}
\caption{Spectrum of Sgr B2(N) from 271.5 - 272.0 GHz}
\end{figure}

\clearpage

\begin{figure}
\plotone{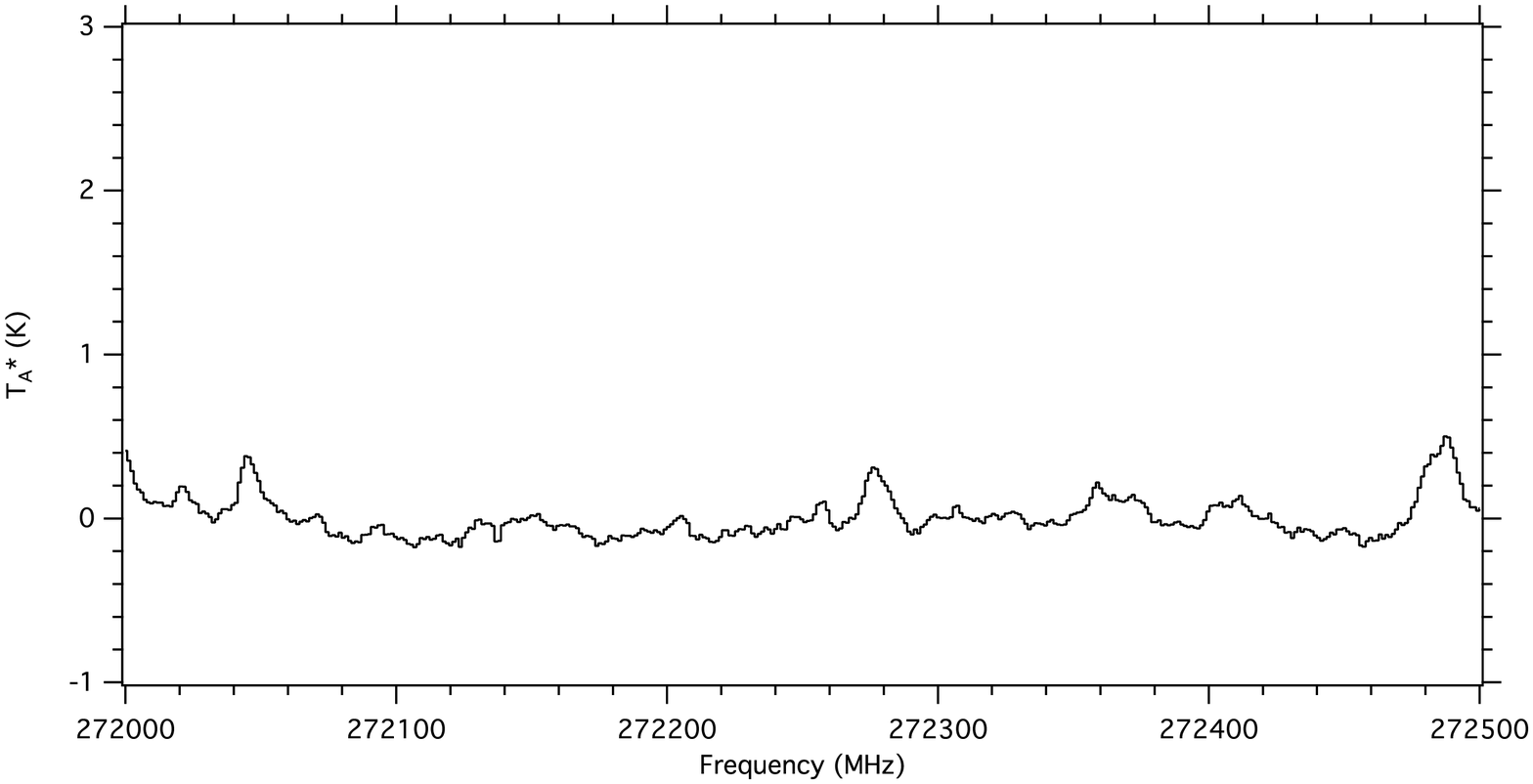}
\caption{Spectrum of Sgr B2(N) from 272.0 - 272.5 GHz}
\end{figure}

\begin{figure}
\plotone{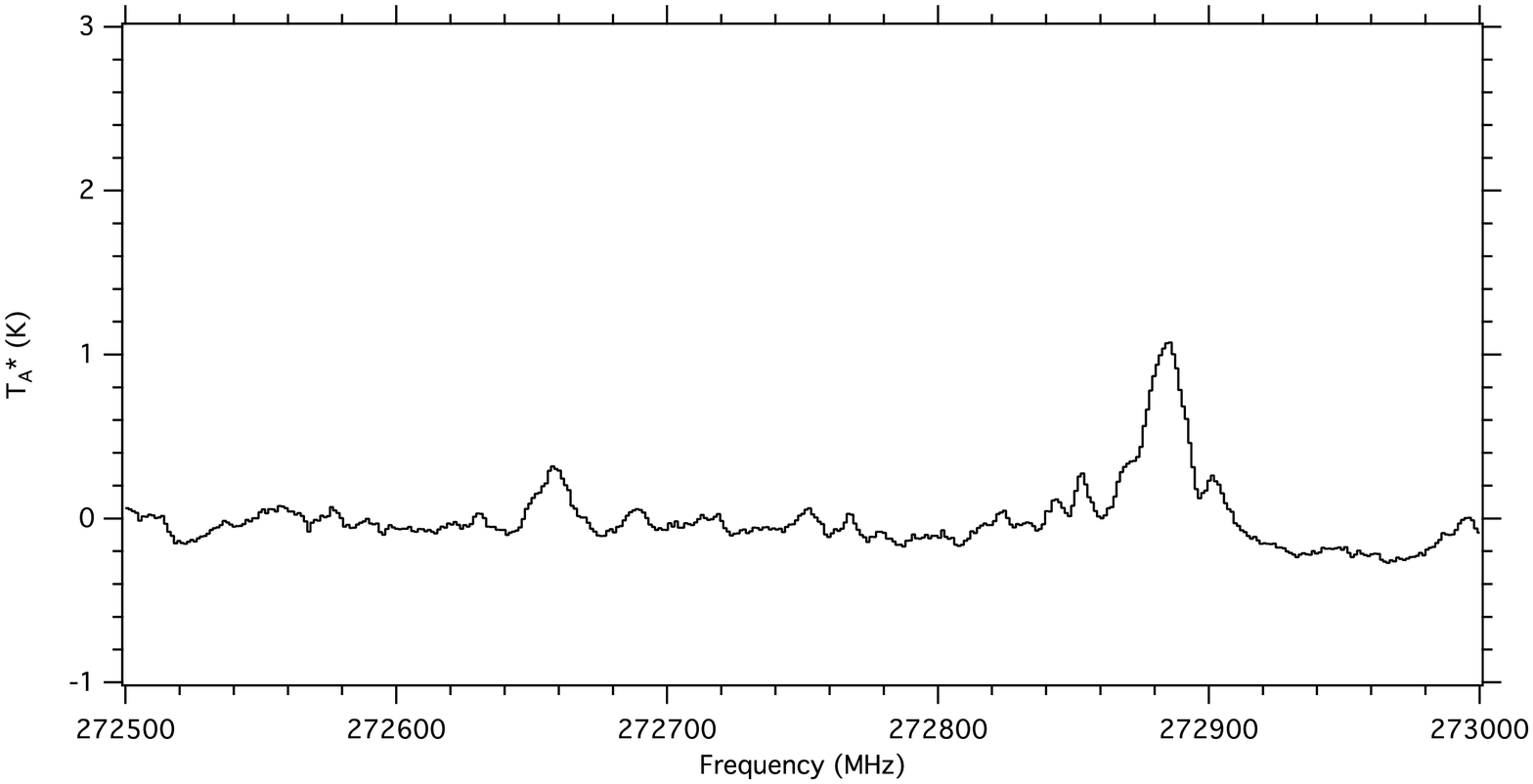}
\caption{Spectrum of Sgr B2(N) from 272.5 - 273.0 GHz}
\end{figure}

\clearpage

\begin{figure}
\plotone{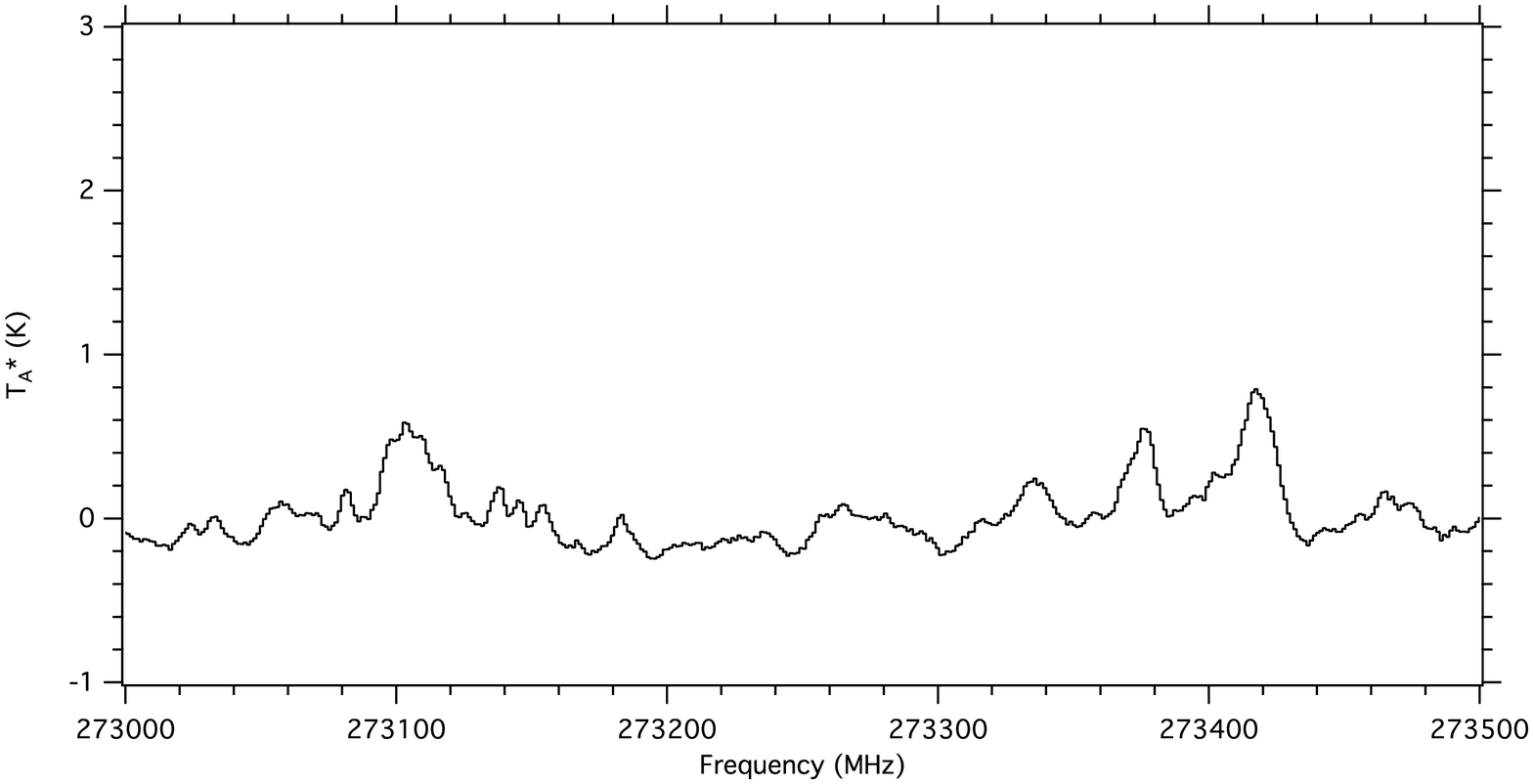}
\caption{Spectrum of Sgr B2(N) from 273.0 - 273.5 GHz}
\end{figure}

\begin{figure}
\plotone{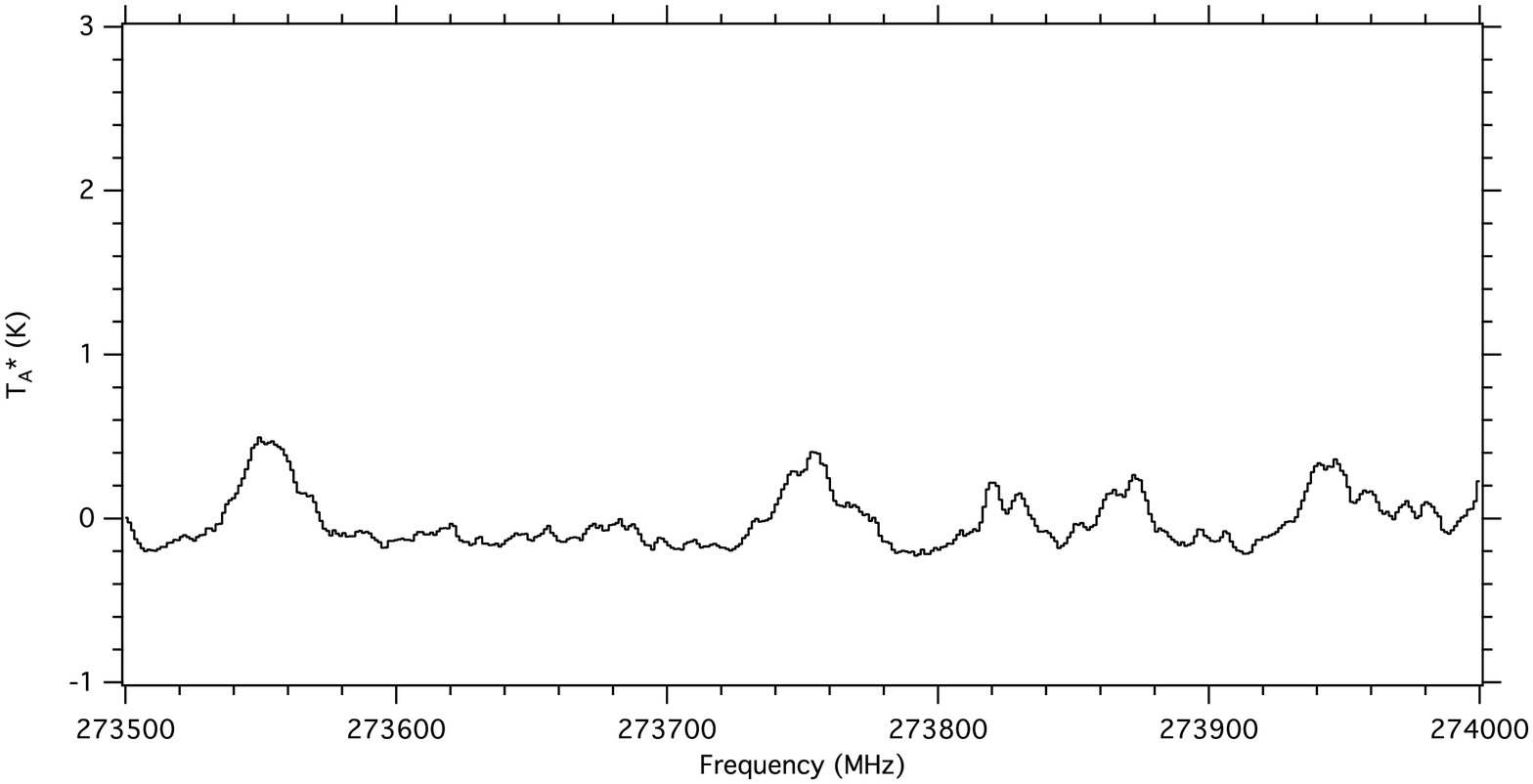}
\caption{Spectrum of Sgr B2(N) from 273.5 - 274.0 GHz}
\end{figure}

\clearpage

\begin{figure}
\plotone{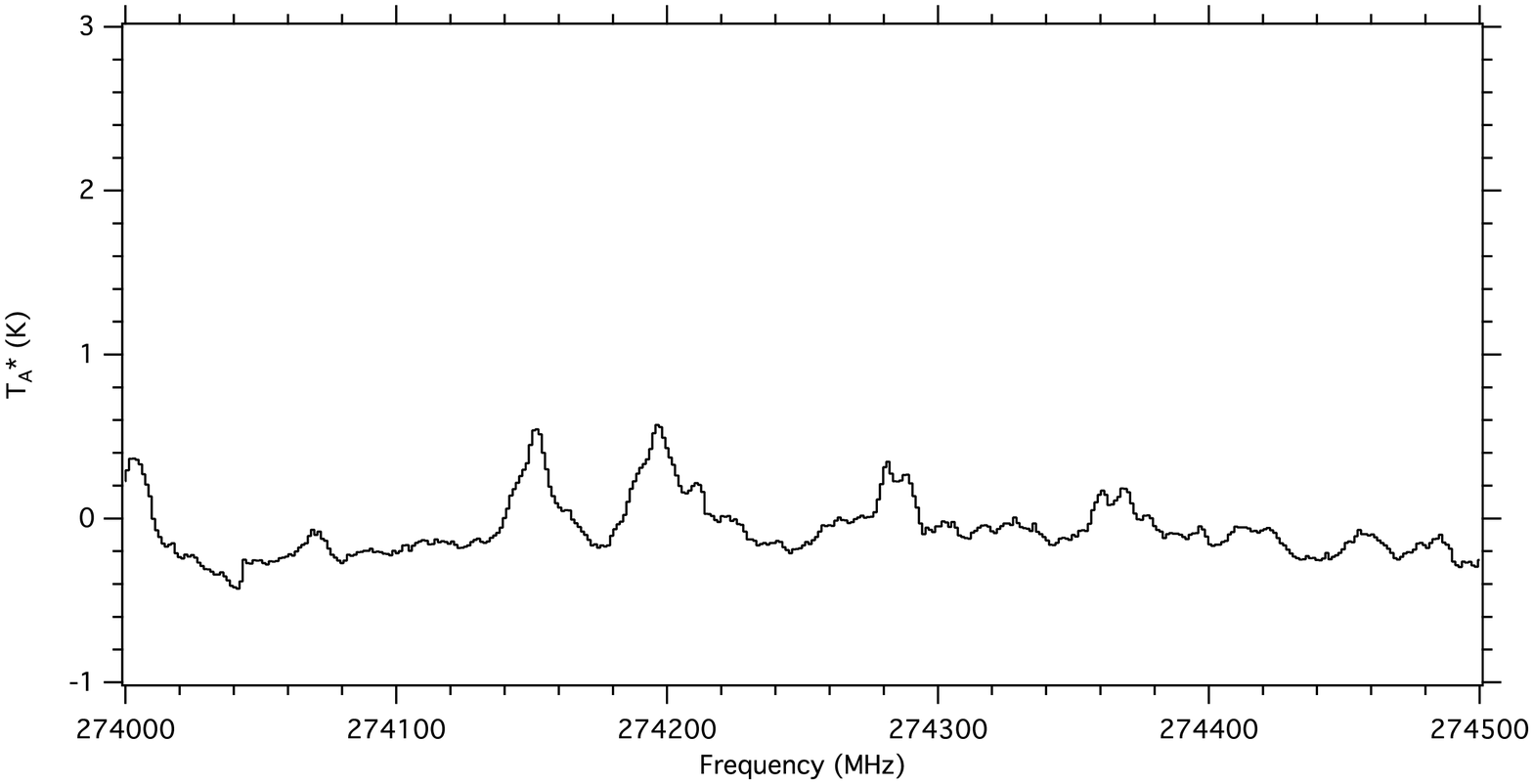}
\caption{Spectrum of Sgr B2(N) from 274.0 - 274.5 GHz}
\end{figure}

\begin{figure}
\plotone{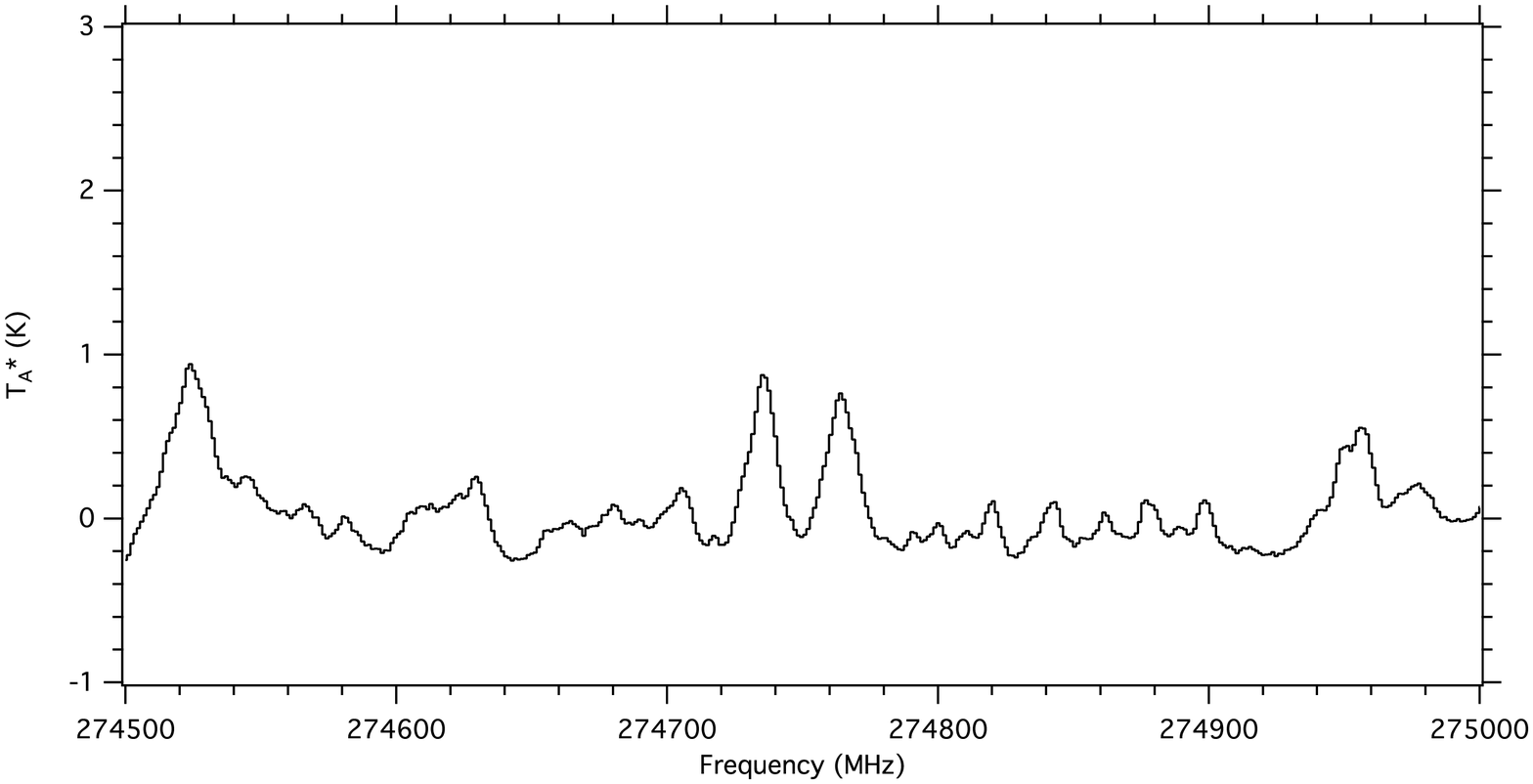}
\caption{Spectrum of Sgr B2(N) from 274.5 - 275.0 GHz}
\end{figure}

\clearpage

\begin{figure}
\plotone{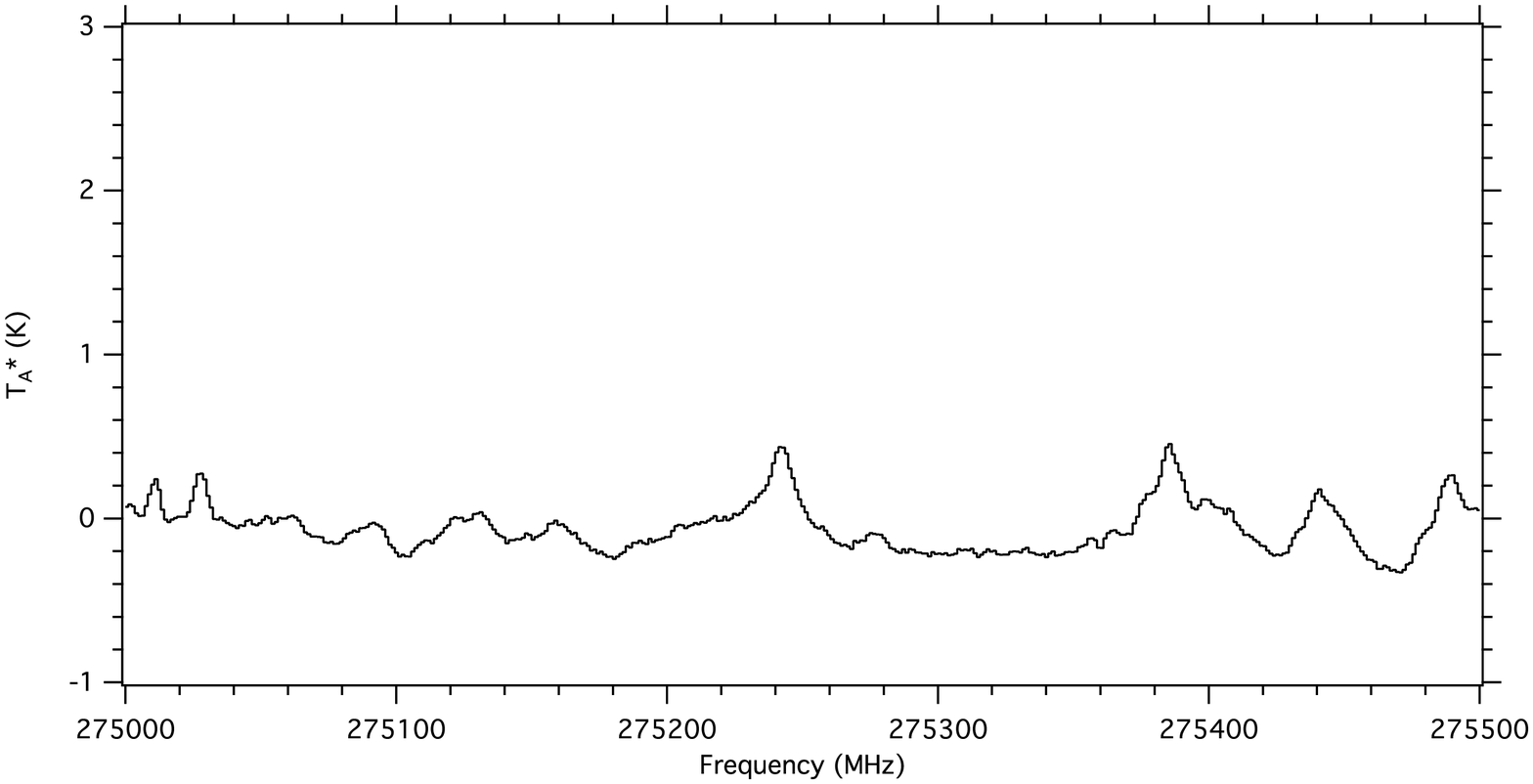}
\caption{Spectrum of Sgr B2(N) from 275.0 - 275.5 GHz}
\end{figure}

\begin{figure}
\plotone{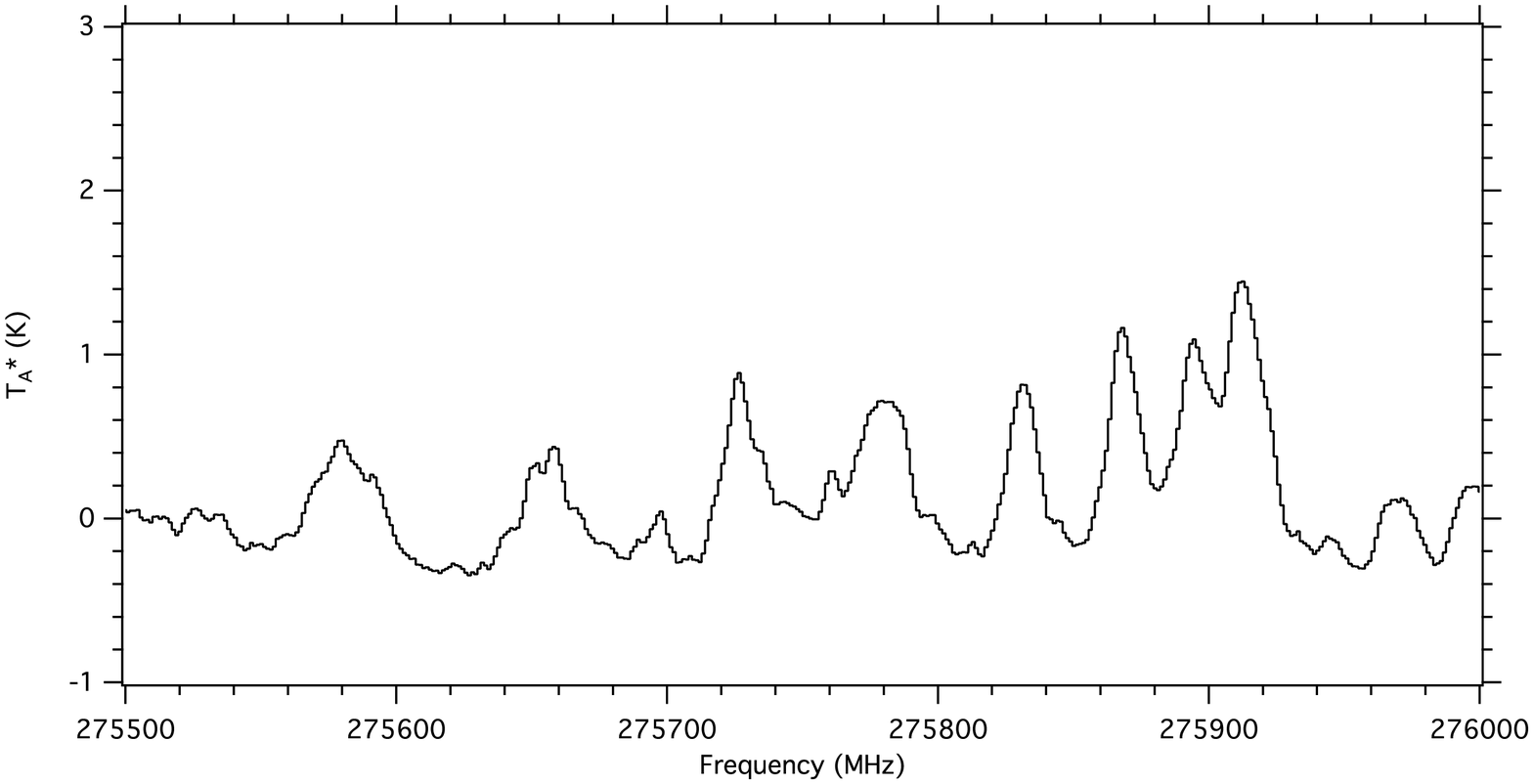}
\caption{Spectrum of Sgr B2(N) from 275.5 - 276.0 GHz}
\end{figure}

\clearpage

\begin{figure}
\plotone{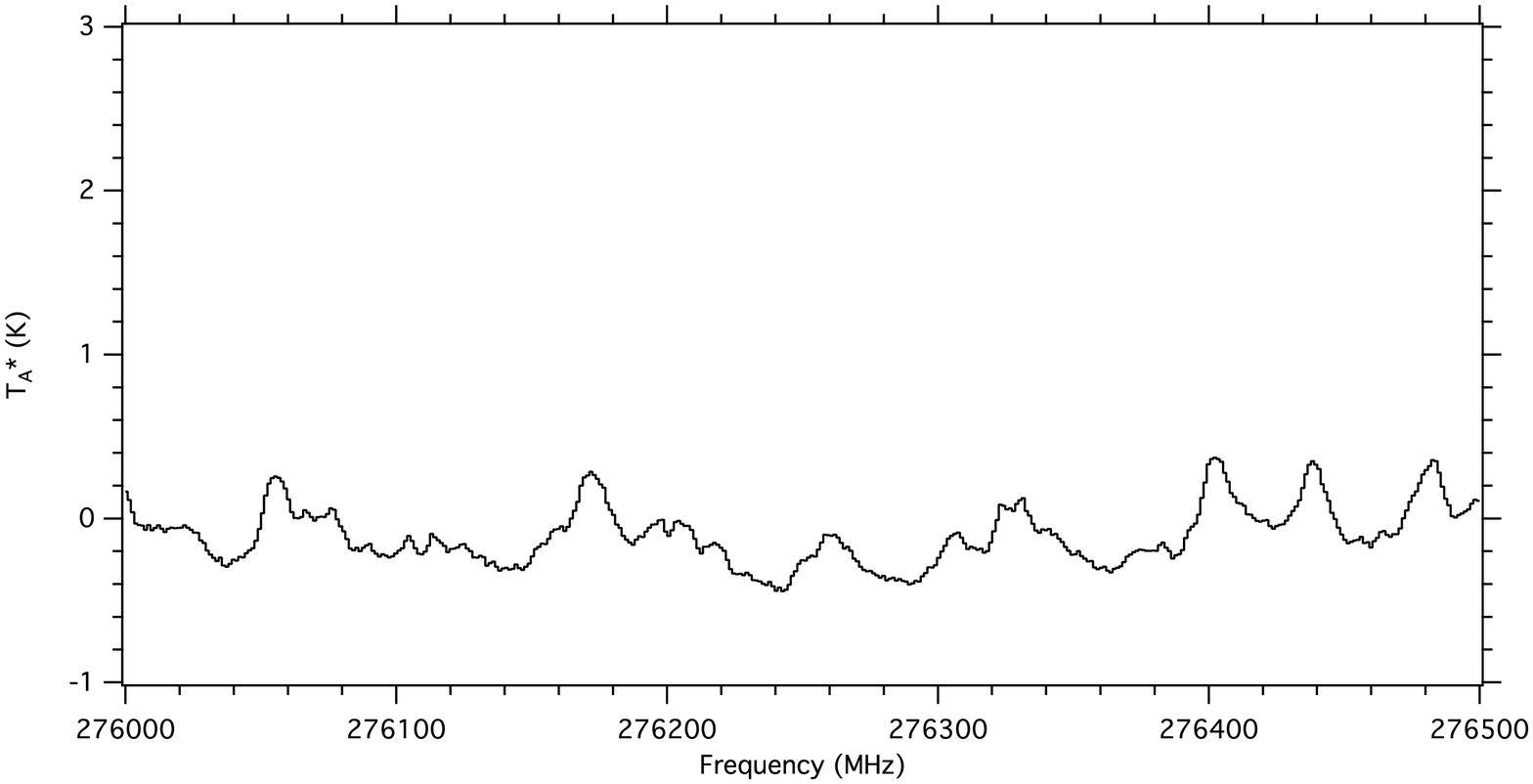}
\caption{Spectrum of Sgr B2(N) from 276.0 - 276.5 GHz}
\end{figure}

\begin{figure}
\plotone{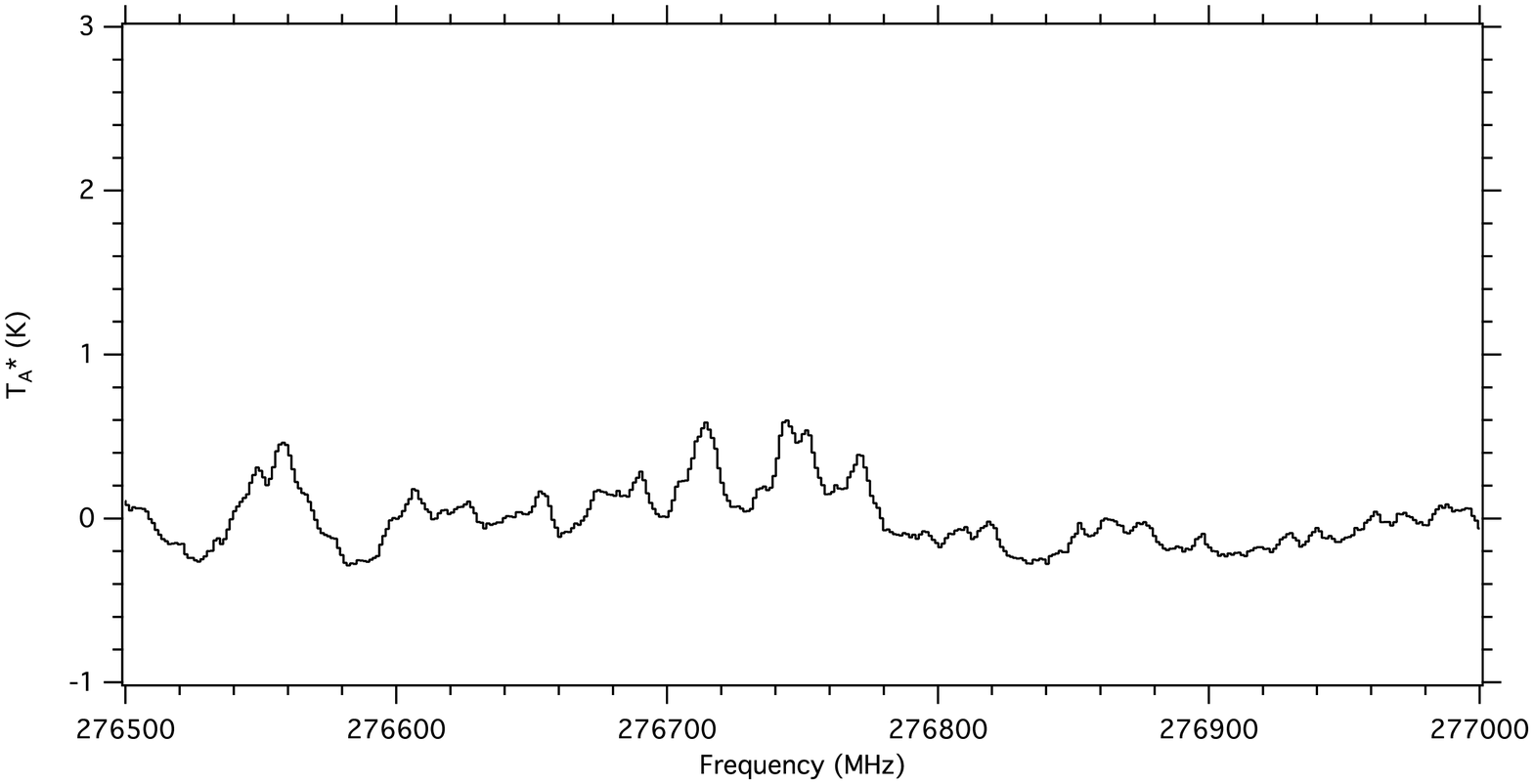}
\caption{Spectrum of Sgr B2(N) from 276.5 - 277.0 GHz}
\end{figure}

\clearpage

\begin{figure}
\plotone{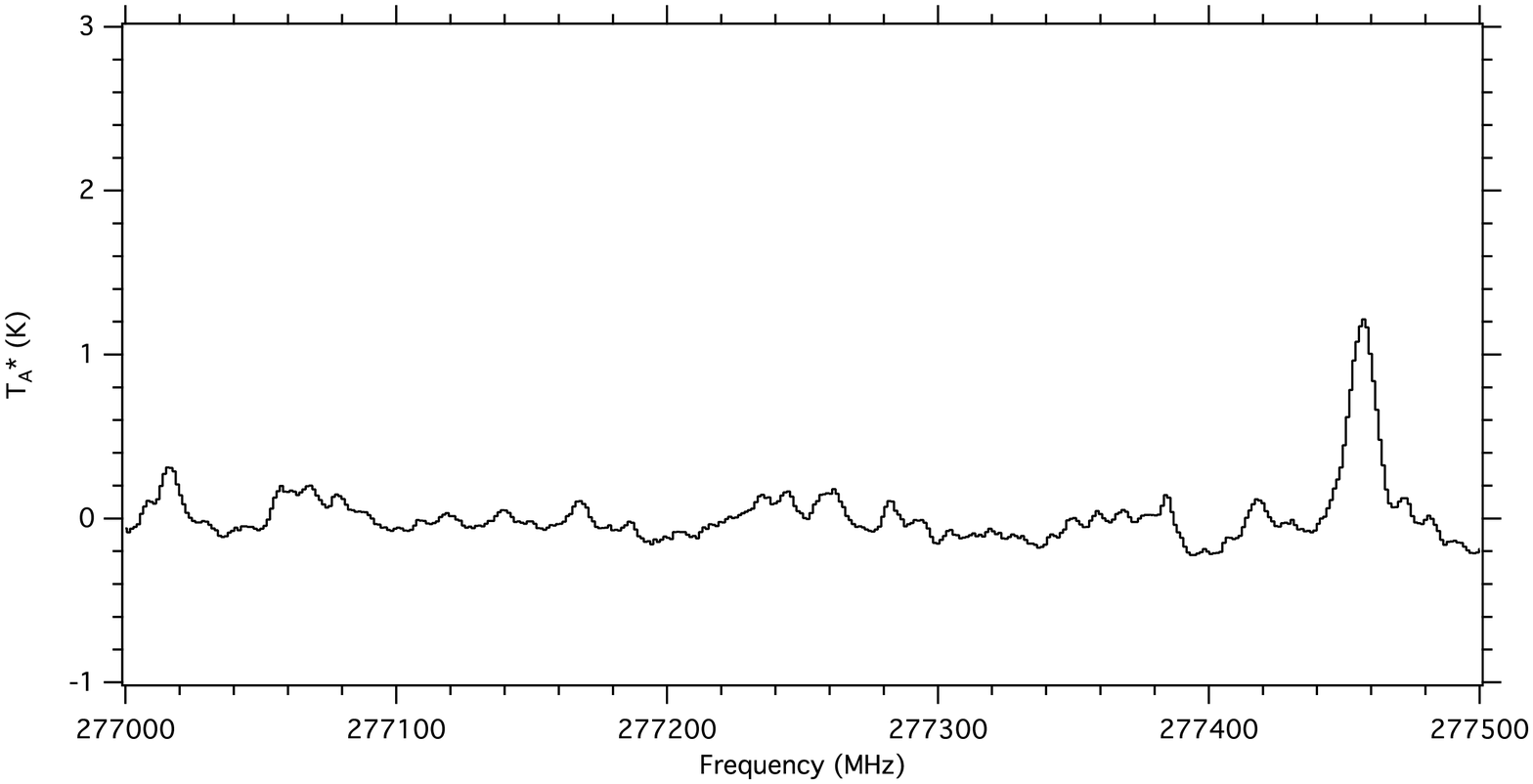}
\caption{Spectrum of Sgr B2(N) from 277.0 - 277.5 GHz}
\end{figure}

\begin{figure}
\plotone{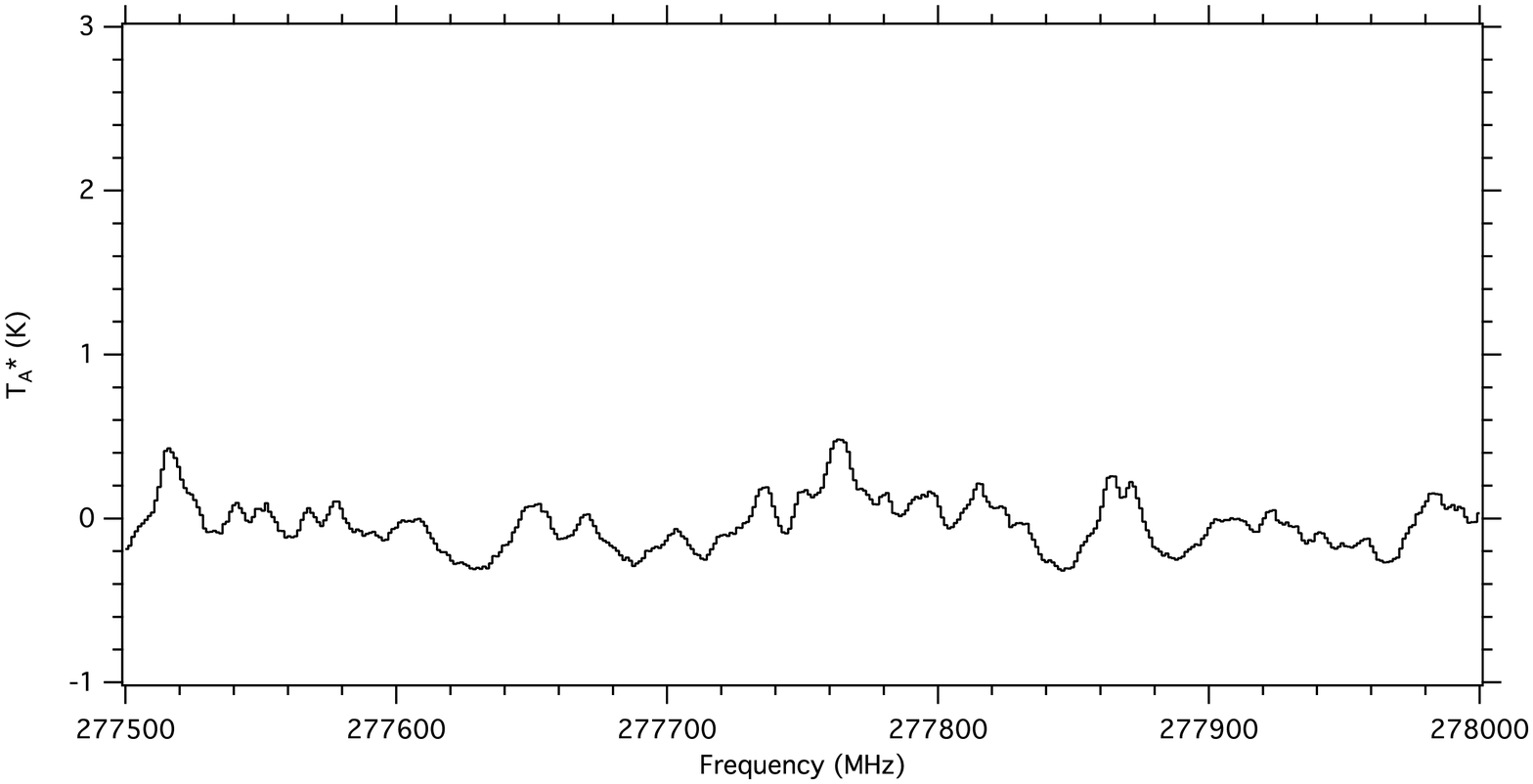}
\caption{Spectrum of Sgr B2(N) from 277.5 - 278.0 GHz}
\end{figure}

\clearpage

\begin{figure}
\plotone{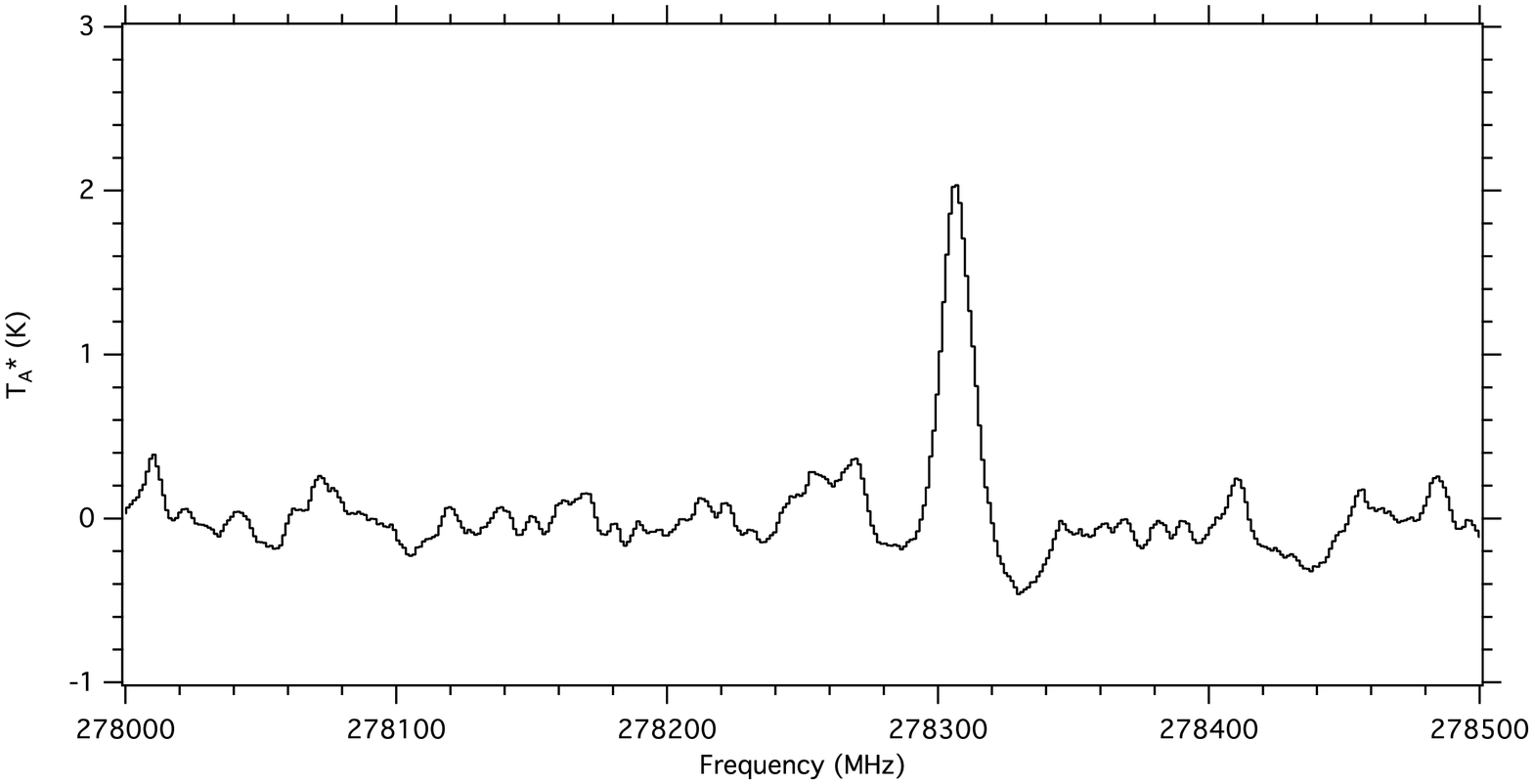}
\caption{Spectrum of Sgr B2(N) from 278.0 - 278.5 GHz}
\end{figure}

\begin{figure}
\plotone{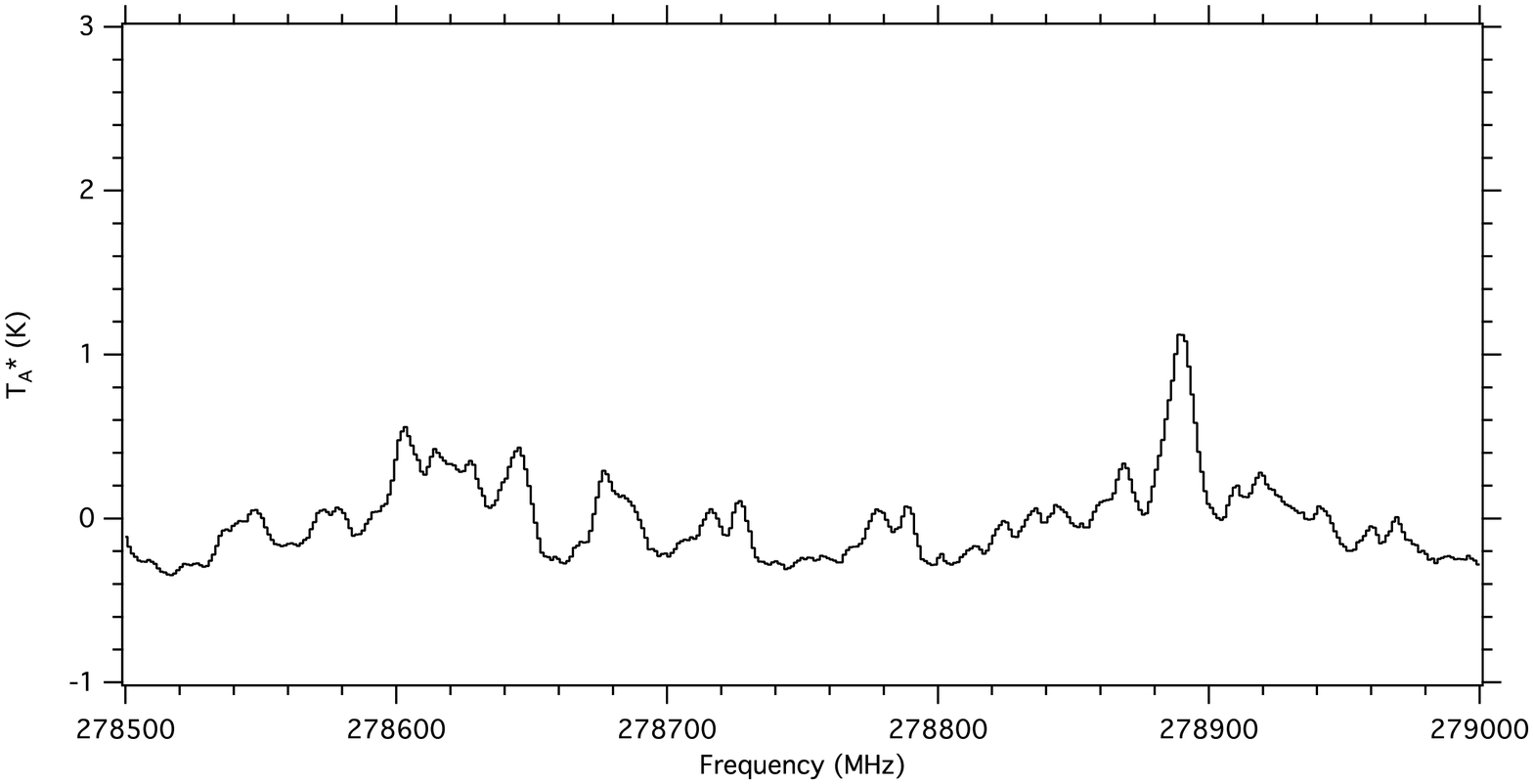}
\caption{Spectrum of Sgr B2(N) from 278.5 - 279.0 GHz}
\end{figure}

\clearpage

\begin{figure}
\plotone{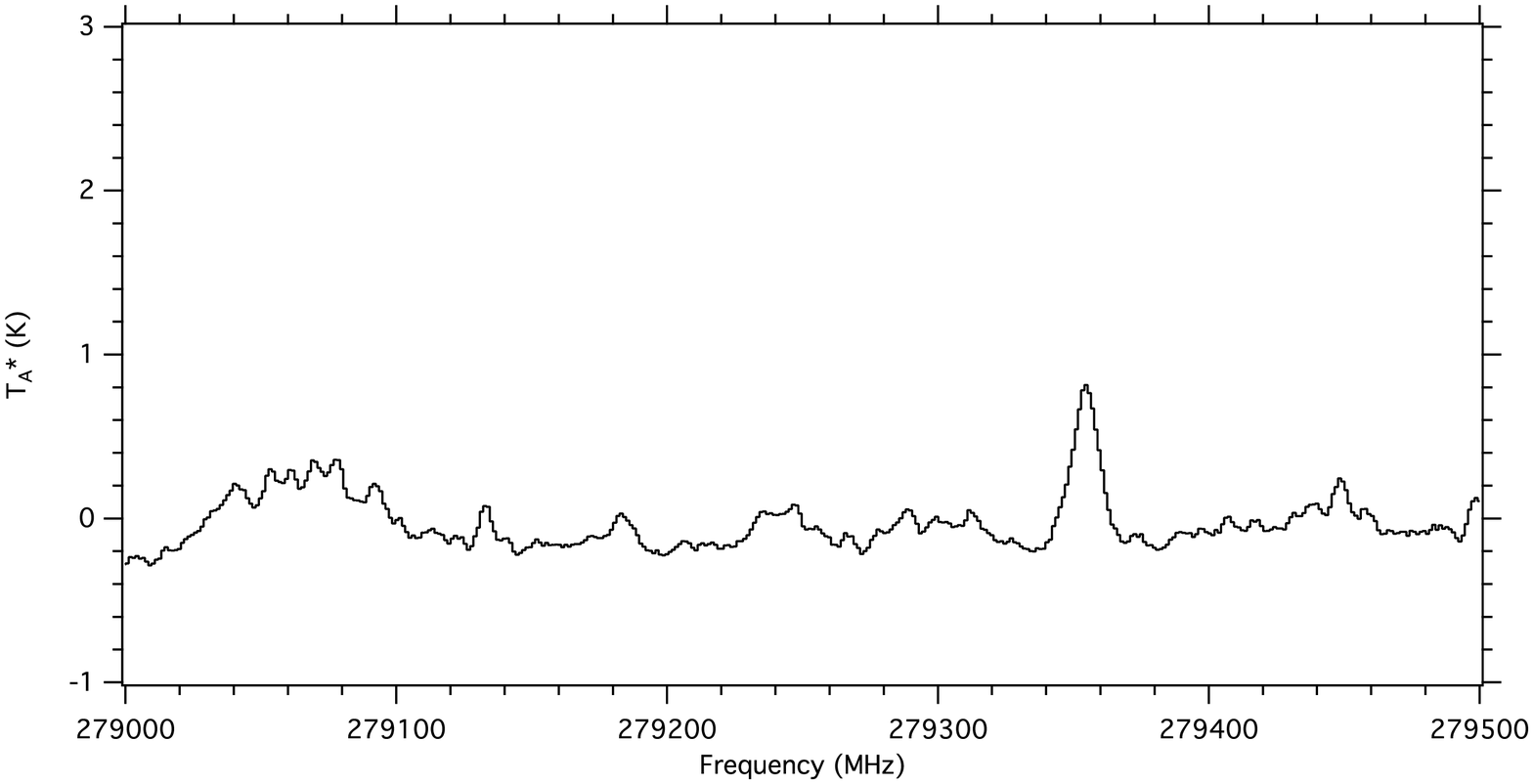}
\caption{Spectrum of Sgr B2(N) from 279.0 - 279.5 GHz}
\end{figure}

\begin{figure}
\plotone{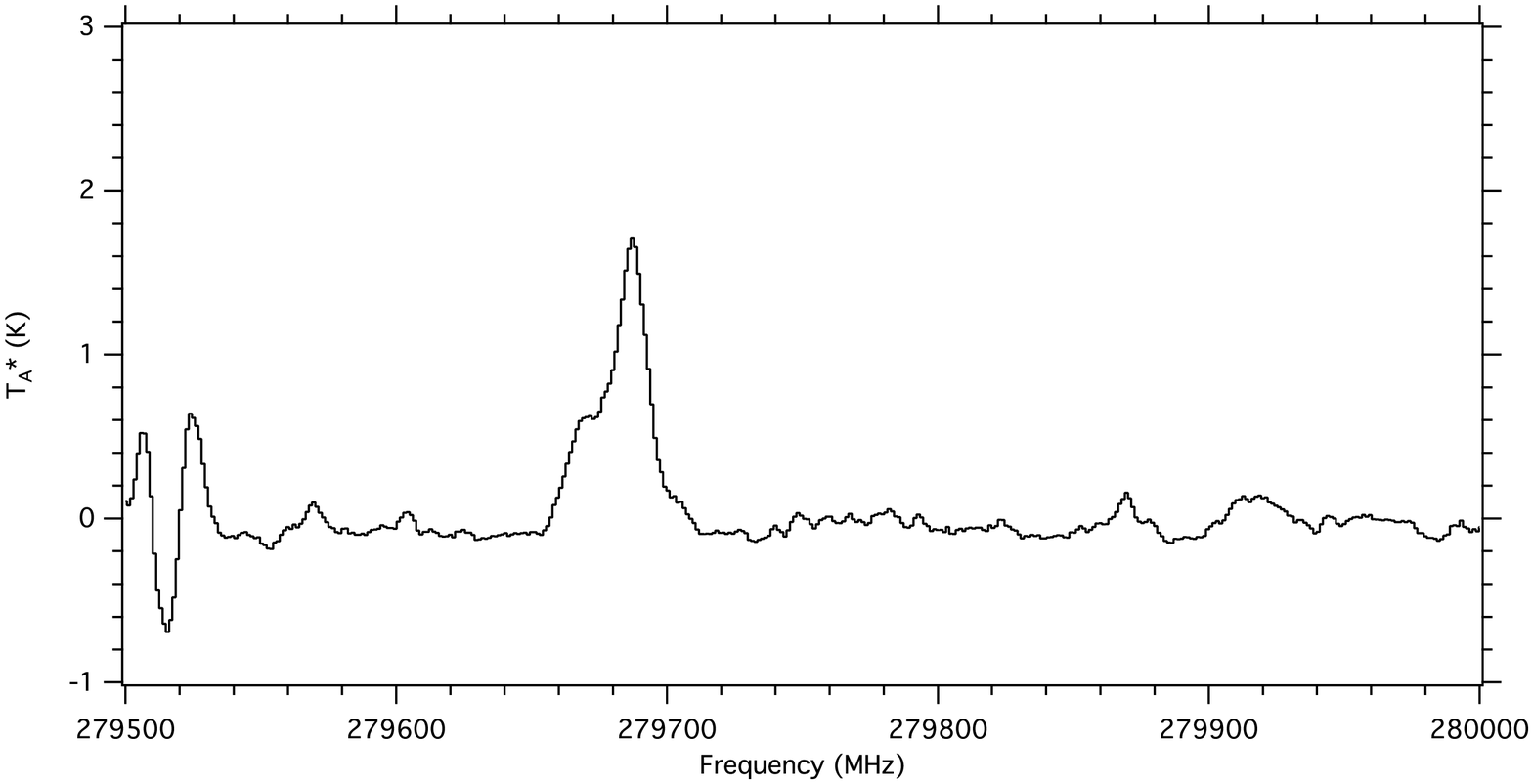}
\caption{Spectrum of Sgr B2(N) from 279.5 - 280.0 GHz}
\end{figure}

\clearpage

\begin{figure}
\plotone{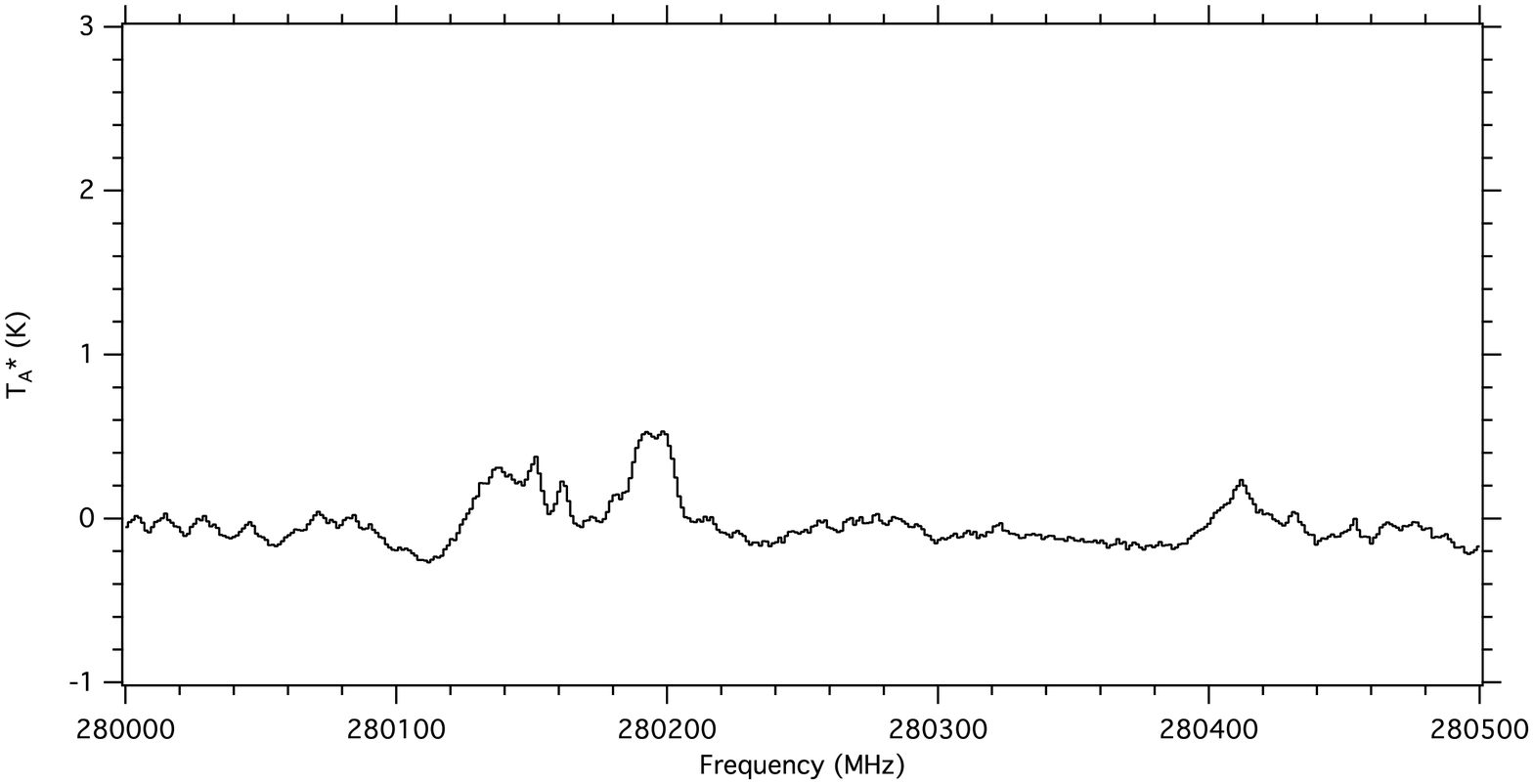}
\caption{Spectrum of Sgr B2(N) from 280.0 - 280.5 GHz}
\end{figure}

\begin{figure}
\plotone{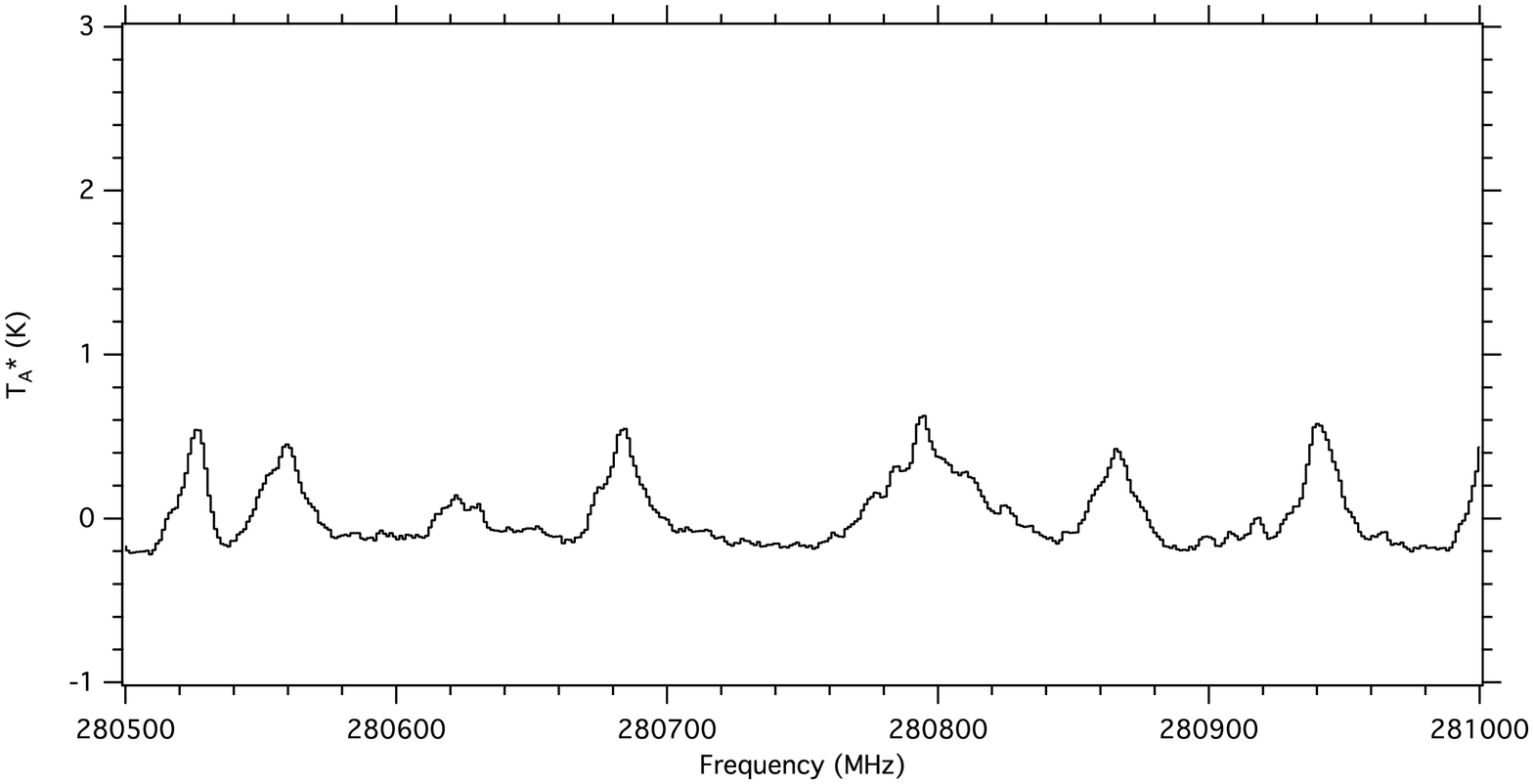}
\caption{Spectrum of Sgr B2(N) from 280.5 - 281.0 GHz}
\end{figure}

\clearpage

\begin{figure}
\plotone{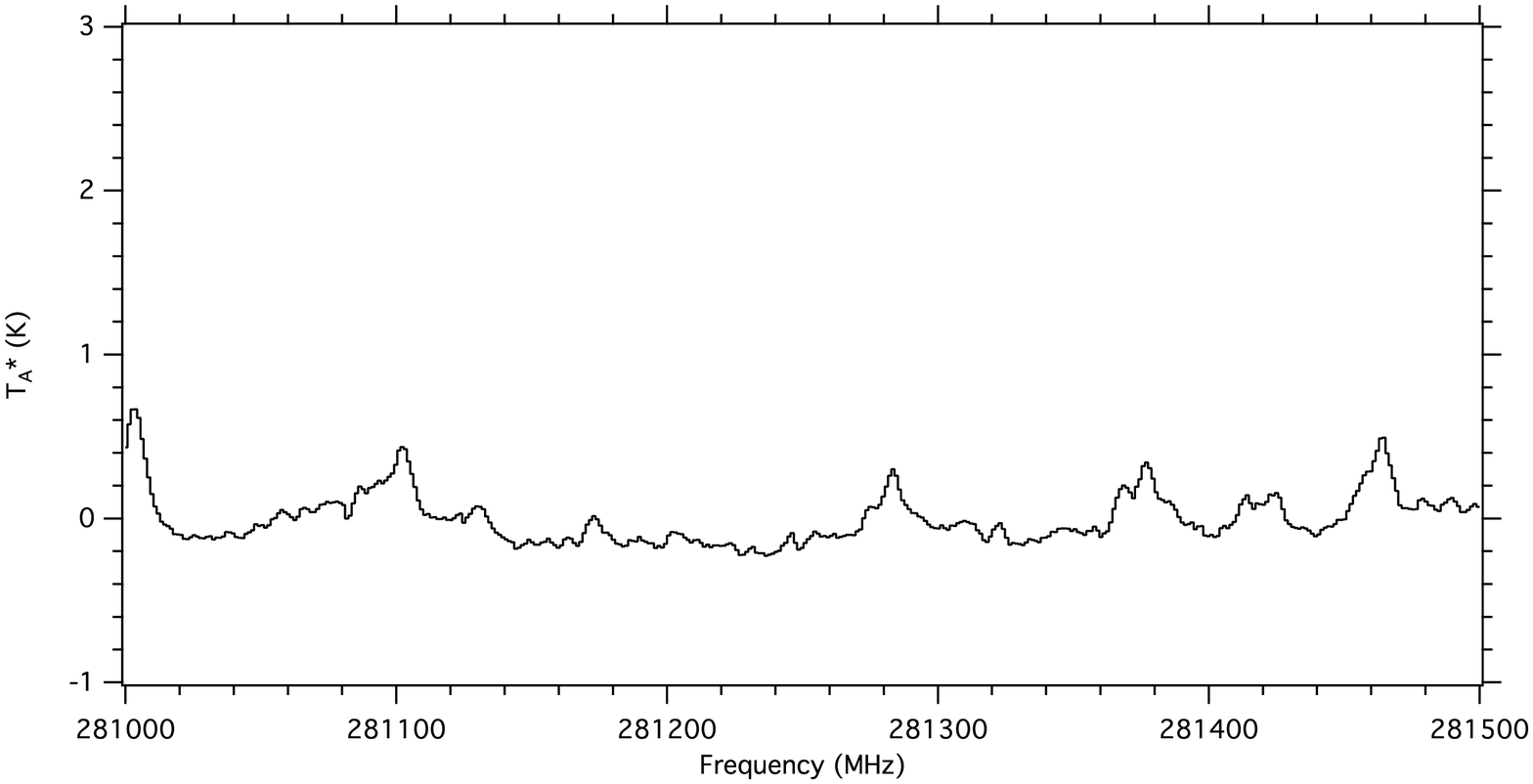}
\caption{Spectrum of Sgr B2(N) from 281.0 - 281.5 GHz}
\end{figure}

\begin{figure}
\plotone{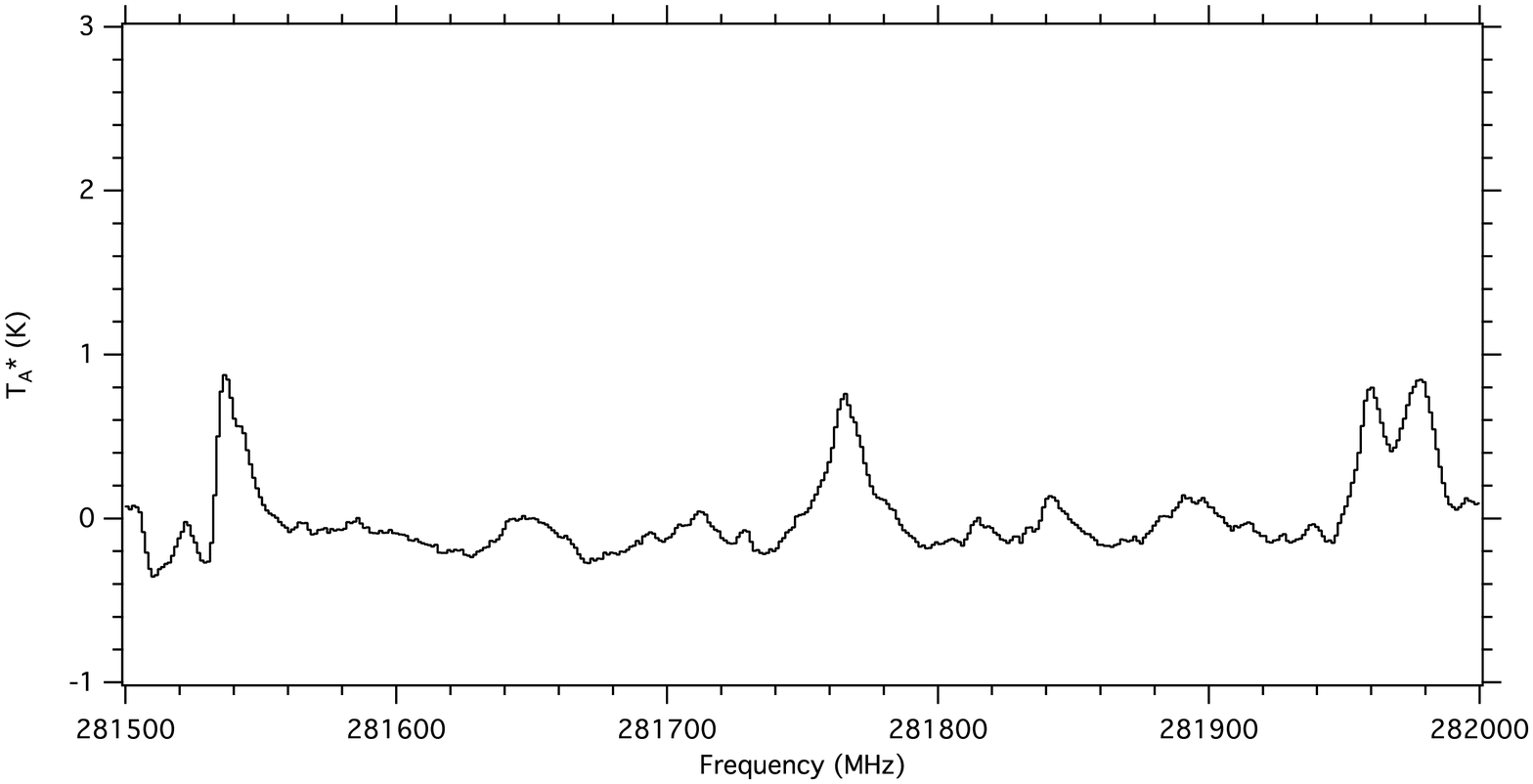}
\caption{Spectrum of Sgr B2(N) from 281.5 - 282.0 GHz}
\end{figure}

\clearpage

\begin{figure}
\plotone{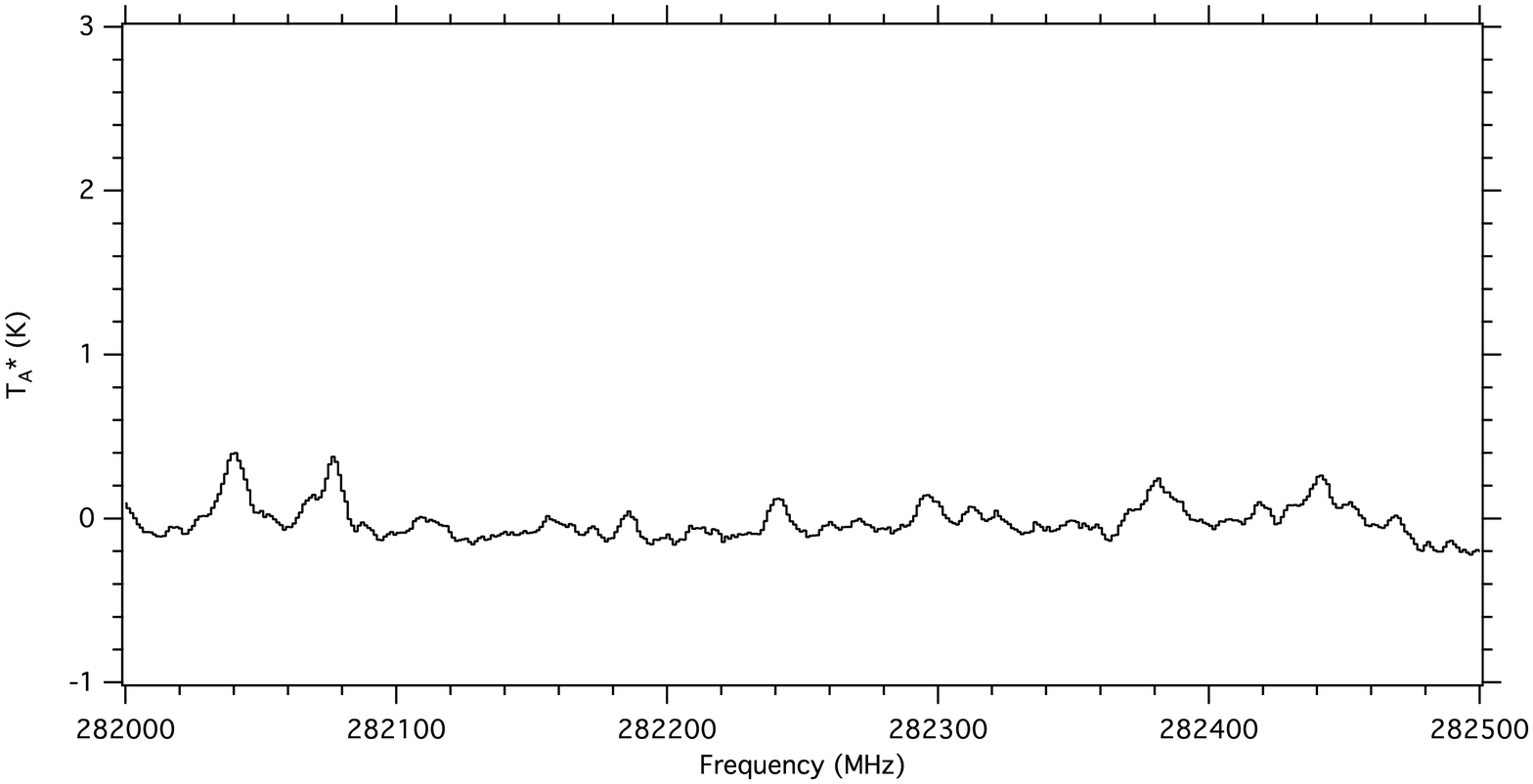}
\caption{Spectrum of Sgr B2(N) from 282.0 - 282.5 GHz}
\end{figure}

\begin{figure}
\plotone{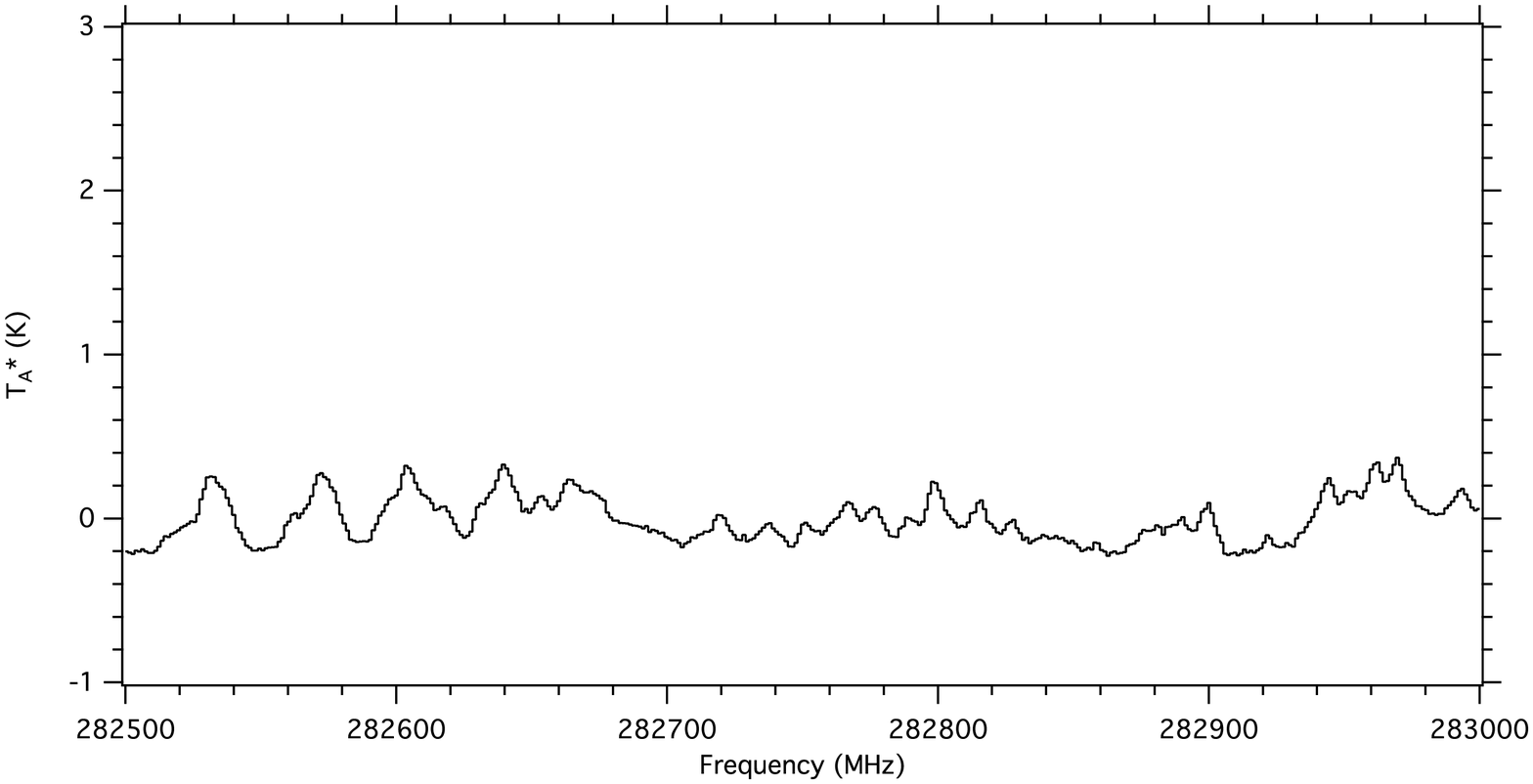}
\caption{Spectrum of Sgr B2(N) from 282.5 - 283.0 GHz}
\end{figure}

\clearpage

\begin{figure}
\plotone{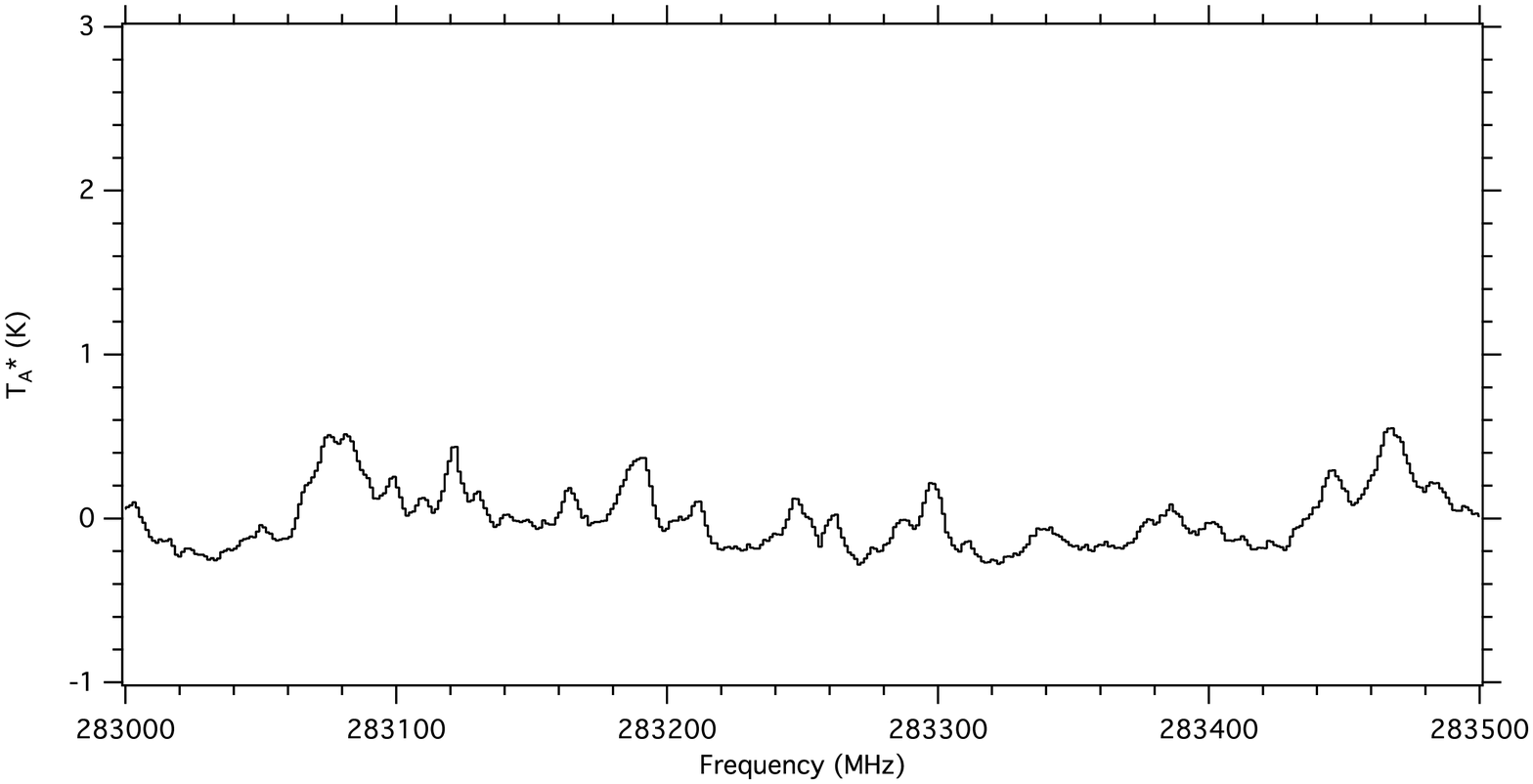}
\caption{Spectrum of Sgr B2(N) from 283.0 - 283.5 GHz}
\end{figure}

\begin{figure}
\plotone{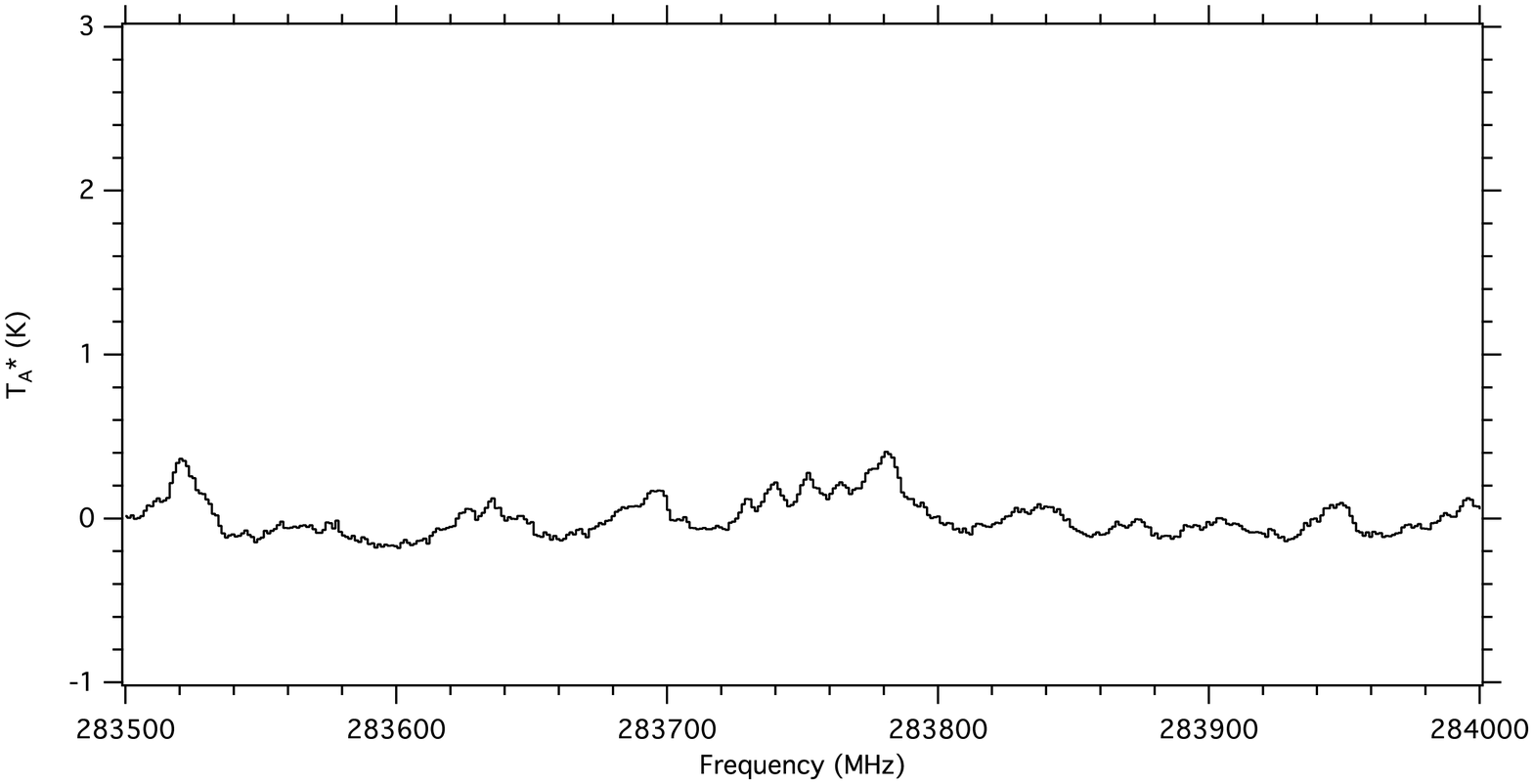}
\caption{Spectrum of Sgr B2(N) from 283.5 - 284.0 GHz}
\end{figure}

\clearpage

\begin{figure}
\plotone{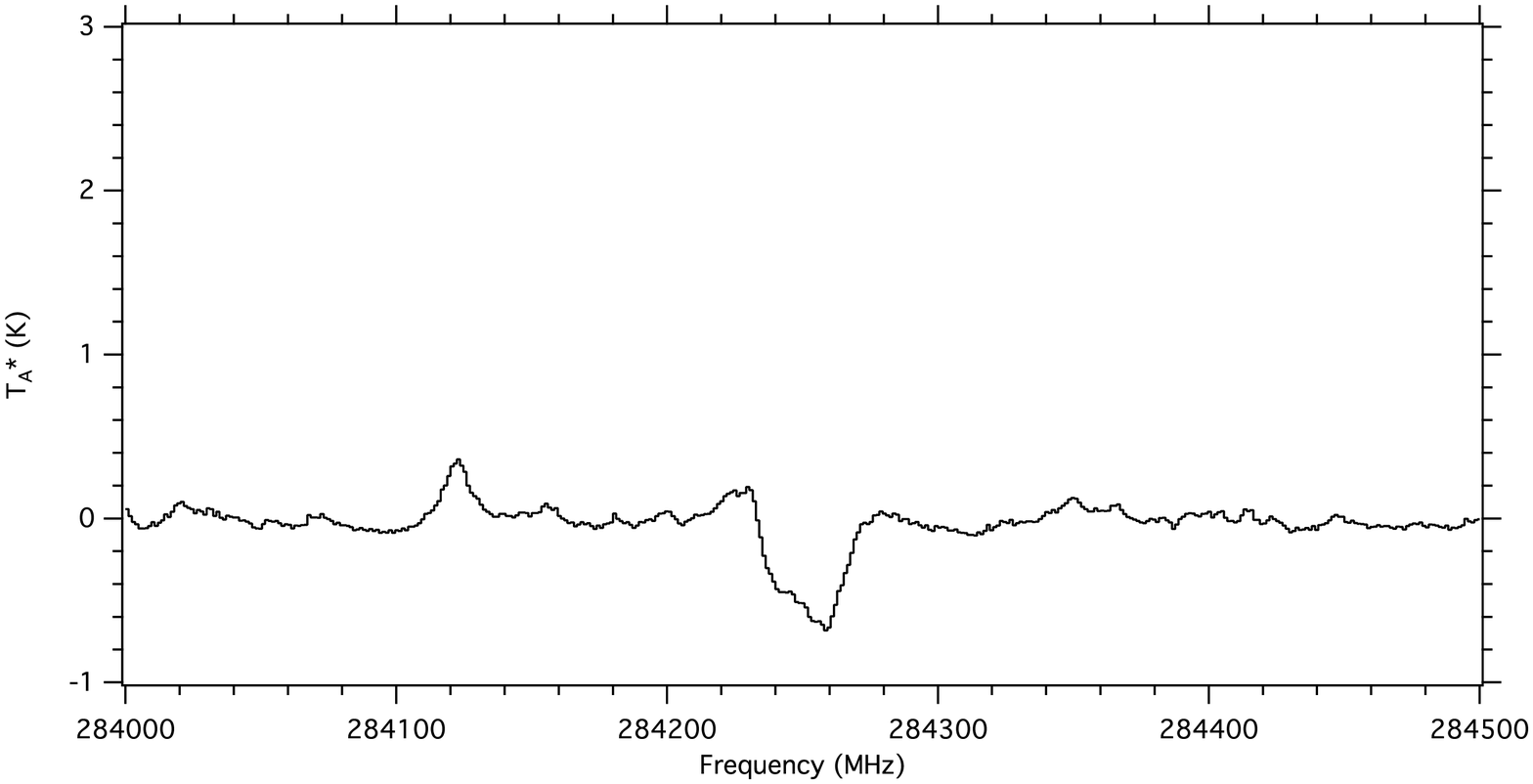}
\caption{Spectrum of Sgr B2(N) from 284.0 - 284.5 GHz}
\end{figure}

\begin{figure}
\plotone{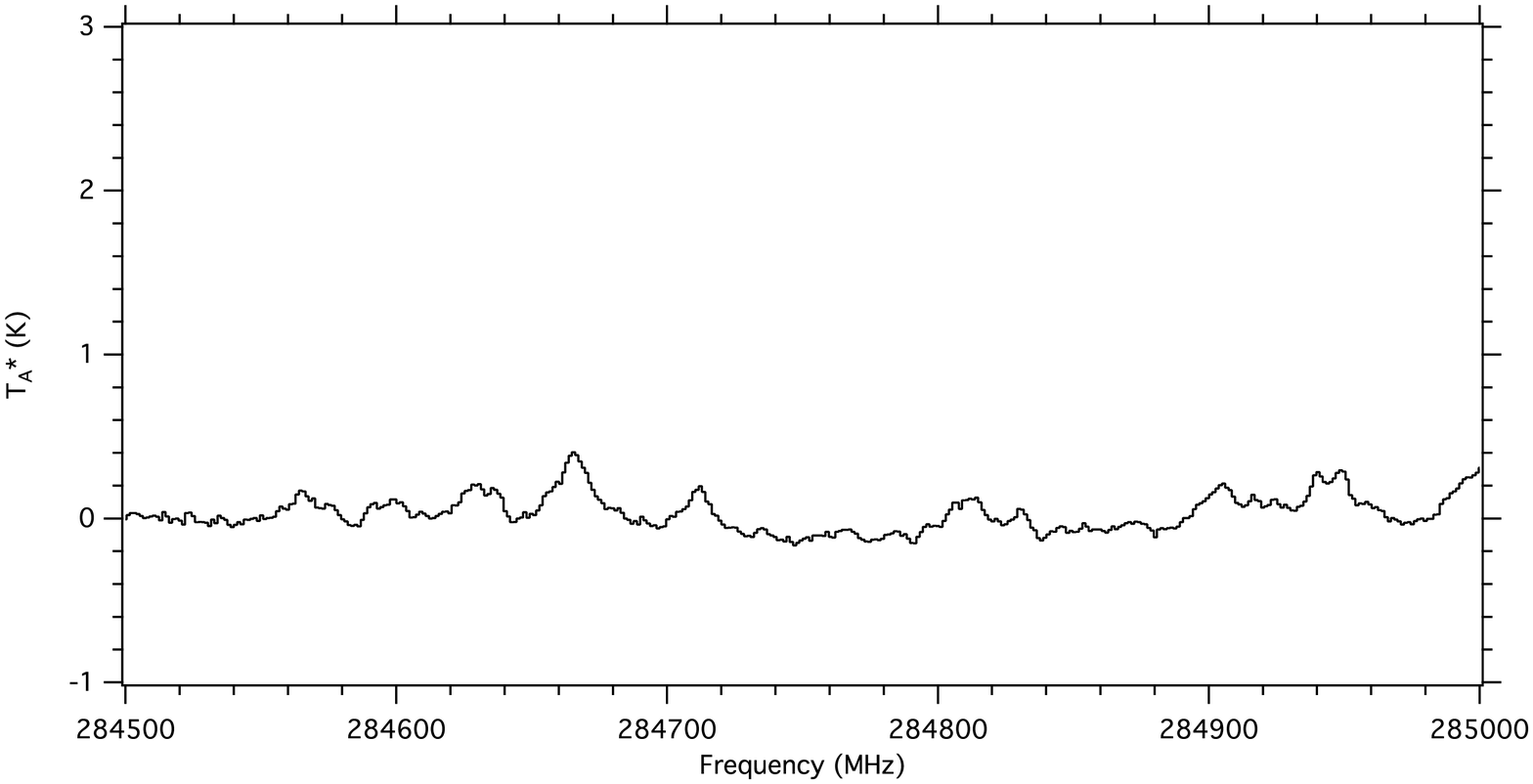}
\caption{Spectrum of Sgr B2(N) from 284.5 - 285.0 GHz}
\end{figure}

\clearpage

\begin{figure}
\plotone{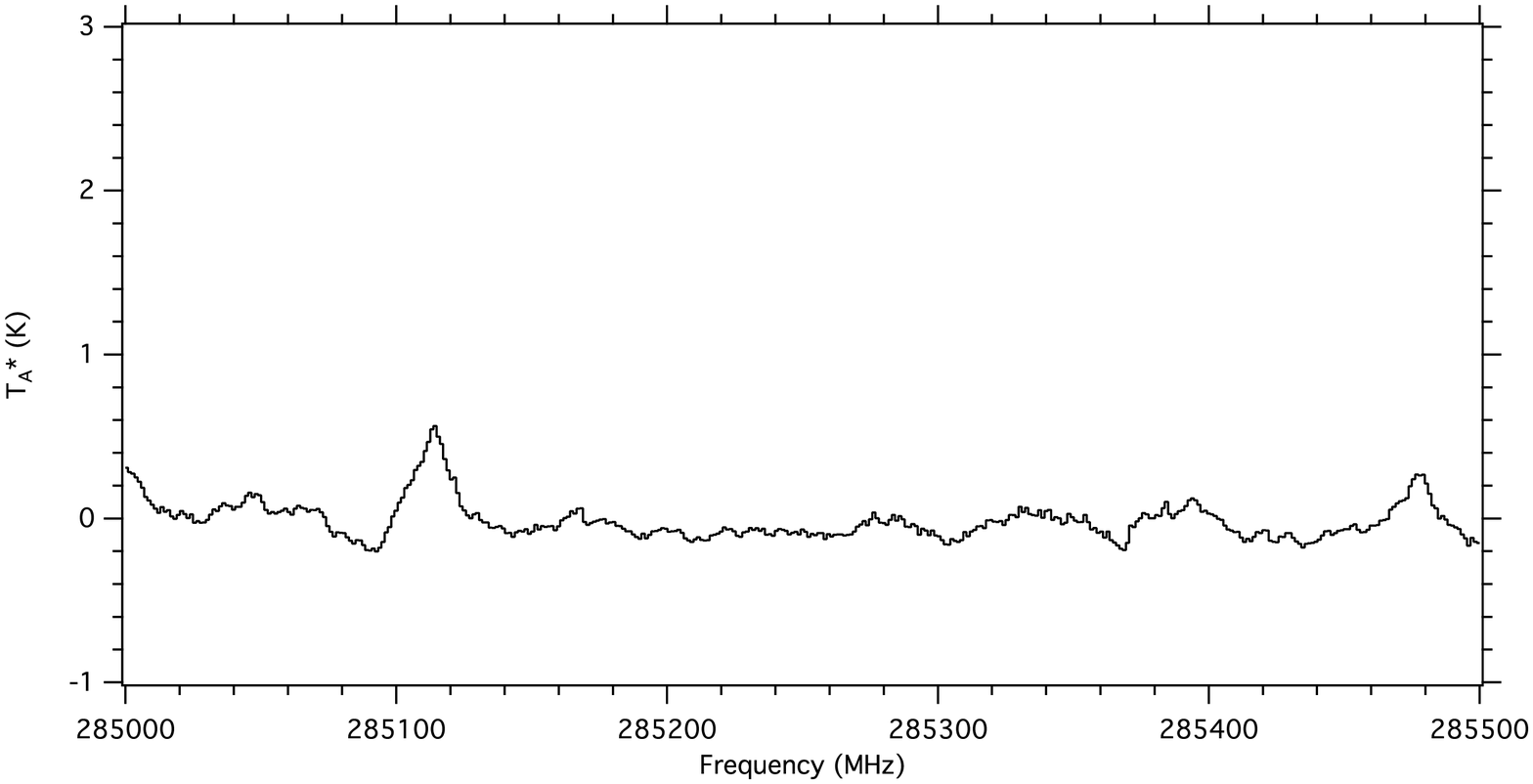}
\caption{Spectrum of Sgr B2(N) from 285.0 - 285.5 GHz}
\end{figure}

\begin{figure}
\plotone{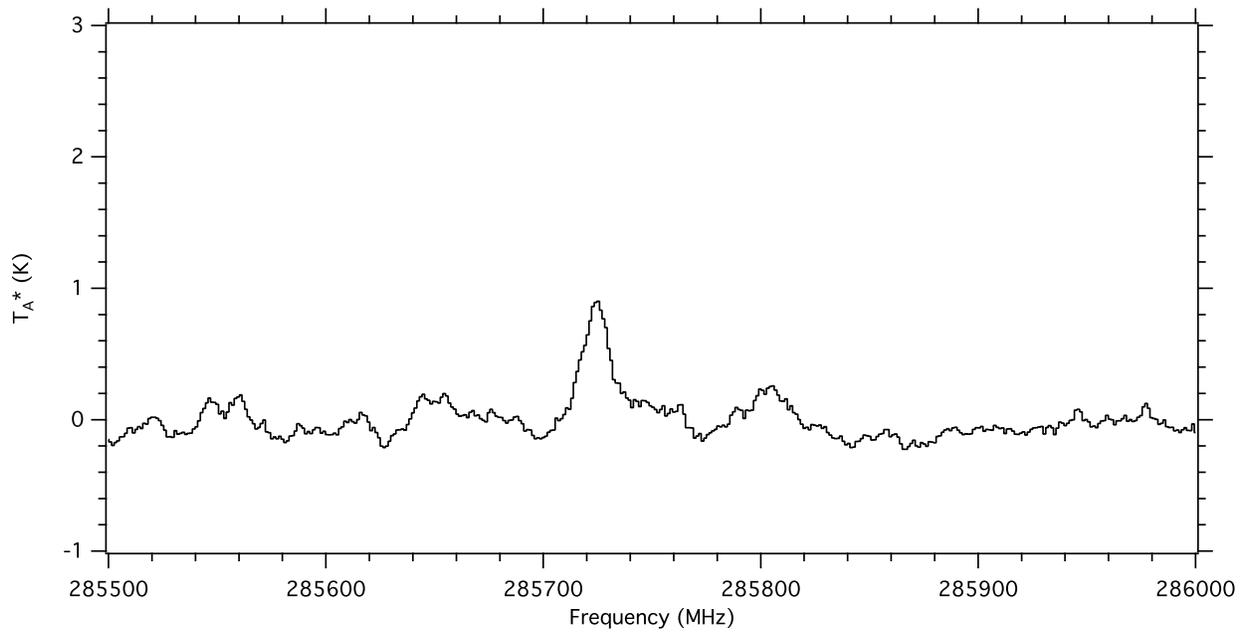}
\caption{Spectrum of Sgr B2(N) from 285.5 - 286.0 GHz}
\end{figure}

\clearpage

\appendix
\section*{Appendix A}

\begin{verbatim}

file in sgrb2n.dat
$rm sgrb2n.sm
file out sgrb2n.sm multiple
set tele "CSO FFTS 1W"
find

for i 1 to found
	get n
	smooth
	write
next

set tele "CSO FFTS 2B1"
find

for i 1 to found
	get n
	write
next

set tele "CSO FFTS 2B2"
find

for i 1 to found
	get n
	write
next

set tele
set mode x t
set mode y t

file in sgrb2n.sm
$rm sgrb2n.rs
file out sgrb2n.rs multiple
set u c
set var spec write

find
for i 1 to found 
	get n 
	a\resam 'nchan/0.3*fres' 'rchan/0.3*fres' int(frequency) 0.3 f /nofft
	write
next

set tele
set mode x t
set mode y t

file in sgrb2n.rs
$rm sgrb2n.cln
file out sgrb2n.cln multiple
set u c 

set tele "CSO FFTS 1W"
find

for i 1 to found
	get n
	write
next

set tele "CSO FFTS 2B1"
find

for i 1 to found
	get n
	for j 3666 to 3670
		draw k j
	next
	for j 5499 to 5503
		draw k j
	next
	for j 1832 to 1837
		draw k j
	next
	write
next

set tele "CSO FFTS 2B2"
find

for i 1 to found
	get n
	for j 3666 to 3670
		draw k j
	next
	for j 5499 to 5503
		draw k j
	next
	for j 1832 to 1837
		draw k j
	next
	for j 5330 to 5337
		draw k j
	next
	write
next

set tele
set mode x t
set mode y t

file in sgrb2n.cln
$rm sgrb2n.bas
file out sgrb2n.bas multiple
find

drop 4211
drop 4206
drop 4242
drop 4060
drop 4229
drop 4150

for i 1 to found
	get n
	set win 0 1
	base 3
	write
next

set tele
set mode x t
set mode y t

$rm sgrb2n.dec
$rm sgrb2n.igor
file in sgrb2n.bas
file out sgrb2n.dec multiple
set unit f f
set mode x t
set blanking -1000 0.1
find
initialize
set var spec read

deconv 1.e-6 400
write 1

smooth
smooth

write 2

file in sgrb2n.dec
find
get 1
plot
let ry 0 /where ry.eq.(-1000)

greg sgrb2n.txt /formatted

get 2
plot
let ry 0 /where ry.eq.(-1000)

greg sgrb2n.sm.txt /formatted

\end{verbatim}

\end{document}